\documentclass{sigplanconf}
\usepackage{hyperref}
\usepackage{breakurl}
\usepackage{times}
\usepackage{shading}
\usepackage{amssymb}
\usepackage{amsmath}
\usepackage{amsthm}
\usepackage{stmaryrd}
\usepackage{mathpartir}
\usepackage[dvips]{graphicx}
\usepackage{color}

\newenvironment{defn}{\begin{tabbing}
  \hspace{1.5em} \= \hspace{.20\linewidth - 1.5em} \= \hspace{1.5em} \= \kill
  }{
  \end{tabbing}}
\newenvironment{defn2}{\begin{tabbing}
  \hspace{1.5em} \= \hspace{.295\linewidth - 1.5em} \= \hspace{1.5em} \= \kill
  }{
  \end{tabbing}}

\newcounter{ctr:list1}
\newcounter{ctr:list2}

\newenvironment{list2'}
    {\setcounter{ctr:list2}{0}
    \begin{list}{(\roman{ctr:list2})}{\usecounter{ctr:list2}\setlength{\leftmargin}{0.3cm}}}
    {\end{list}}

\makeatletter 
\def\overbracket#1{\mathop{\vbox{\ialign{##\crcr\noalign{\kern3\p@} \downbracketfill\crcr\noalign{\kern3\p@\nointerlineskip} $\hfil\displaystyle{#1}\hfil$\crcr}}}\limits} 
\def\underbracket#1{\mathop{\vtop{\ialign{##\crcr $\hfil\displaystyle{#1}\hfil$\crcr\noalign{\kern3\p@\nointerlineskip} \upbracketfill\crcr\noalign{\kern3\p@}}}}\limits} \def\overparenthesis#1{\mathop{\vbox{\ialign{##\crcr\noalign{\kern3\p@} \downparenthfill\crcr\noalign{\kern3\p@\nointerlineskip} $\hfil\displaystyle{#1}\hfil$\crcr}}}\limits} 
\def\underparenthesis#1{\mathop{\vtop{\ialign{##\crcr
	$\hfil\displaystyle{#1}\hfil$\crcr\noalign{\kern3\p@\nointerlineskip} \upparenthfill\crcr\noalign{\kern3\p@}}}}\limits} 
\def\downparenthfill{$\m@th\braceld\leaders\vrule\hfill\bracerd$} 
\def\upparenthfill{$\m@th\bracelu\leaders\vrule\hfill\braceru$} 
\def\upbracketfill{$\m@th\makesm@sh{\llap{\vrule\@height3\p@\@width.7\p@}}%
\leaders\vrule\@height.7\p@\hfill
\makesm@sh{\rlap{\vrule\@height3\p@\@width.7\p@}}$} 
\def\downbracketfill{$\m@th \makesm@sh{\llap{\vrule\@height.7\p@\@depth2.3\p@\@width.7\p@}}%
\leaders\vrule\@height.7\p@\hfill \makesm@sh{\rlap{\vrule\@height.7\p@\@depth2.3\p@\@width.7\p@}}$}
\def\equalsfill{$\m@th\mathord=\mkern-7mu \cleaders\hbox{$\!\mathord=\!$}\hfill \mkern-7mu\mathord=$}  
\makeatother 

\newcommand{\cenv}[3]{
\begin{flushleft}
\parbox{8.3cm}{{\bf #1} $~#2$}
\\
\vspace{-0.1cm}
\parbox{8.3cm}{\downbracketfill}
\\
\vspace{-0.3cm}
\end{flushleft}
#3
\begin{flushleft}
\vspace{-0.2cm}
\parbox{8.3cm}{\upbracketfill}
\end{flushleft}}

\newcommand{\cenvvv}[3]{
\vspace{0.8mm}
\begin{flushleft}
\parbox{8.4cm}{{\bf #1} $~#2$}
\\
\parbox{8.4cm}{\downbracketfill}
\\
\vspace{-0.2cm}
\end{flushleft}
#3
\begin{flushleft}
\parbox{8.4cm}{\upbracketfill}
\end{flushleft}}

\newcommand{\entry}[2]{\>$#1$\>\>#2}
\newcommand{\clause}[2]{$#1$\>\>#2}
\newcommand{\mycategory}[2]{\clause{#1::=}{#2}}

\newcommand{\lab}{\mathsf L}
\newcommand{\labp}{\mathsf P}
\newcommand{\labo}{\mathsf O}
\newcommand{\labb}{\mathsf S}
\newcommand{\labt}{\mathsf E}
\newcommand{\labc}{\mathsf C}

\newcommand{\new}[2]{(\nu #1)\:#2}

\newcommand{\action}[1]{\stackrel{#1}{\longrightarrow}~}
\newcommand{\fn}{\mathtt{fn}}
\newcommand{\bn}{\mathtt{bn}}
\newcommand{\fv}{\mathtt{fv}}
\newcommand{\bv}{\mathtt{bv}}

\newcommand{\fork}[2]{#1\Rsh\:\!#2}
\newcommand{\eval}[3]{\mathsf{let}~#1=#2~\mathsf{in}~#3}
\newcommand{\store}[1]{\stackrel{#1}\mapsto}
\newcommand{\lctx}[2]{\mathcal E_\lab[\![#1]\!]_{#2}}
\newcommand{\llctx}[3]{\mathcal E_{#1}\llbracket#2\rrbracket_{#3}}

\newcommand{\actsub}[1]{\stackrel{#1;\sigma}{\longrightarrow}~}
\newcommand{\sctx}[2]{\mathcal E_{\lab;\sigma}\llbracket#1\rrbracket_{#2}}
\newcommand{\sctxr}[2]{\mathcal E_{\labp;\sigma}\llbracket#1\rrbracket_{#2}}

\newcommand{\actioni}[1]{\stackrel{#1}{\rightarrowtriangle}~}

\newcommand{\trule}[1]{(\textbf{Typ #1})}
\newcommand{\srule}[1]{\textbf{Struct #1}}
\newcommand{\rrule}[1]{\textbf{Reduct #1}}

\newcommand{\dom}{\mathtt{dom}}

\newcommand{\rem}[1]{$\textshade[.91]{sharpcorners}{\gdef\outlineboxwidth{0.01}$#1$}$}

\newenvironment{compact}
        {\begin{list}{$\bullet$}{
        }}
        {\end{list}}

\newcounter{compactenumc}

\newenvironment{compactenum2}
        {\begin{list}{(\roman{compactenumc})}{
        \usecounter{compactenumc}
        \setlength{\leftmargin}{7mm}
        \setlength{\labelwidth}{\leftmargin}
        }}
        {\end{list}}

\newenvironment{compactenum3}
        {\begin{list}{(\arabic{compactenumc})}{
        \usecounter{compactenumc}
        \setlength{\leftmargin}{6mm}
        \setlength{\labelwidth}{\leftmargin}
        }}
        {\end{list}}

\newtheorem{definition}{Definition}[section]
\newtheorem{theorem}[definition]{Theorem}
\newtheorem{lemma}[definition]{Lemma}

\newtheorem{proposition}[definition]{Proposition}

\newtheorem{example}[definition]{Example}

\title{A Type System for Data-Flow Integrity on Windows Vista}
\authorinfo
        {Avik Chaudhuri}
        {University of California at Santa Cruz}
        {avik@cs.ucsc.edu}
\authorinfo
        {Prasad Naldurg \and Sriram Rajamani}
        {Microsoft Research India}
        {\{prasadn,sriram\}@microsoft.com}

\begin{document}
\conferenceinfo{PLAS'08,} {June 8, 2008, Tucson, Arizona, USA.}
\CopyrightYear{2008}
\copyrightdata{978-1-59593-936-4/08/06}
\maketitle



\begin{abstract} 
%

The Windows Vista operating system implements an interesting model of multi-level integrity.  
We observe that in this model, trusted code can be blamed for any information-flow attack; thus, it is possible to eliminate such attacks by static analysis of trusted code. We formalize this model by designing a type system 
that can efficiently enforce data-flow integrity on Windows Vista. Typechecking guarantees that objects whose
contents are statically trusted never contain untrusted values, regardless of
what untrusted code runs in the environment. 
Some of Windows Vista's runtime access checks are necessary for soundness; others are redundant and can be optimized away. 
\end{abstract}





\section{Introduction}\label{intro}

Commercial operating systems are seldom designed to prevent information-flow attacks. Not surprisingly, such attacks are the source of many serious security problems in these systems \cite{sabelfeld}. Microsoft's Windows Vista operating system implements an integrity model that can potentially prevent such attacks. In some ways, this model resembles other, classical models of multi-level integrity~\cite{biba}---every process and object\footnote{In this context, an object may be a file, a channel, a memory location, or indeed any reference to data or executable code.} 
 is tagged with an integrity label, the labels are ordered by levels of trust, and access control is enforced across trust boundaries. In other ways, it is radically different. 
While Windows Vista's access control prevents
low-integrity processes from writing to high-integrity objects, it does not prevent high-integrity processes from reading low-integrity objects.
Further, Windows Vista's integrity labels are
dynamic---labels of processes and objects can change at runtime. This model allows
processes at different trust levels to communicate, and allows dynamic access control. At the same time, it admits various information-flow attacks. Fortunately, it turns out that such attacks require the participation of trusted processes, and can be eliminated by code analysis.

In this paper, we provide a formalization of Windows Vista's integrity model. In particular, we specify an information-flow property called {\em data-flow integrity} (DFI), and present a static type
system that can enforce DFI on Windows Vista. Roughly, DFI prevents any flow of data from the environment to
objects whose contents are trusted. Our type system relies on Windows Vista's
runtime access checks for soundness. The
key idea in the type system is to 
maintain a static lower-bound label~$\labb$ for each object. While the dynamic label of an object can change at runtime, the type system ensures that it
never goes below~$\labb$, and the object never contains a value that
flows from a label lower than~$\labb$.  The label $\labb$ is
declared by the programmer. 
Typechecking requires no other annotations, and can be mechanized by an efficient algorithm.


By design, DFI does not prevent implicit flows \cite{denningcert}. Thus DFI is weaker than noninterference~\cite{nonintf}.
Unfortunately, it is difficult to enforce noninterference on a commercial operating system such as Windows Vista. Implicit flows abound in such systems. Such
flows arise out of frequent, necessary interactions between trusted code
and the environment. They also arise out of covert control channels
which, given the scope of such systems, are impossible to model
sufficiently. Instead, DFI focuses on explicit flows~\cite{denningcert}. This focus buys a reasonable compromise---DFI prevents a definite class of attacks, and can be enforced efficiently on Windows Vista. 
Several successful tools for malware detection follow this approach  \cite{castro,panorama,suh,vogt,dytan,perl}, and a similar approach guides the design of some recent operating systems~\cite{asbestos,histar}. 
Our definition of DFI is dual to standard definitions of
secrecy based on explicit flows---while secrecy prevents sensitive values from
flowing to the environment, DFI prevents the flow of values from the environment to sensitive objects.  Since
there is a rich literature on type-based and logic-based analysis for such definitions of secrecy~\cite{secgp,sectyplog,runtimeprin,ChaudhuriConcur06}, it
makes sense to adapt this analysis for DFI. 
Such an adaptation works, but requires some care.
Unlike secrecy, DFI cannot be enforced without runtime checks. In particular, access checks play a crucial role by
restricting untrusted processes that may run in the environment.  
Further, while secrecy prevents any flow of high-security information to the environment,  
DFI allows certain flows of low-security information from the environment. 
We need to introduce new technical devices for this purpose, including a technique based on \emph{explicit
substitution}~\cite{abadi90explicit} to track precise sources of values. This device is required not only to
specify DFI precisely but also to prove that our type system
enforces DFI.

We design a simple higher-order process calculus that simulates Windows Vista's security environment \cite{bppvista,symantec,uac}. In this language,  
processes can fork new processes, create new objects, change the labels of processes and objects, and read, write, and execute objects in exactly the same ways as Windows Vista allows. 
Our type system exploits Windows Vista's runtime access checks to enforce DFI, and can recognize many correct programs. At the same time, our type system subsumes Windows Vista's execution controls, allowing them to be optimized away.
%

\subsection{Summary of contributions} To sum up, we make the following main contributions in this paper:
\begin{compact}
\item
We propose DFI as a practical multi-level integrity property in the setting of Windows Vista, and formalize DFI using a semantic technique based on explicit substitution.
\item
We present a type system that can efficiently enforce DFI on Windows Vista. Typechecking guarantees DFI regardless 
of what untrusted code runs in the environment. 
\item
We show that while most of Windows Vista's runtime access checks are required to enforce DFI, Windows Vista's execution controls are redundant and can be optimized away.
\end{compact}

\subsection{Outline} The rest of this paper is organized as follows. 
In Section~\ref{overview},
we introduce Windows Vista's security environment, and show how DFI may be violated in that environment. 
In Section~\ref{calc}, we design a calculus that simulates Windows Vista's security environment, equip the calculus with a semantics based on explicit substitution, and formalize DFI in the calculus. 
In Section~\ref{typ}, we
present a system of integrity types and effects for this calculus.  In
Section~\ref{Results}, we prove soundness and other properties of typing. 
Finally, in Section~\ref{concl}, we
discuss limitations and contributions with respect to related work and
conclude. 
Supplementary material, including 
proof details and an efficient typechecking algorithm, appear in the appendix. 
\section{Windows Vista's integrity model}\label{overview} 

%
In this section, we provide a brief overview of Windows Vista's integrity model.\footnote{Windows Vista further implements a discretionary access control model, which we ignore in this paper.} In particular, we introduce Windows Vista's security environment, and show how DFI may be violated in that environment. We observe that such attacks require the participation of trusted processes.

\subsection{Windows Vista's security environment} In Windows Vista, every process and object is tagged with a dynamic integrity label. We indicate such labels in
brackets $(\_)$ below. Labels are related by a total order $\sqsubseteq$, meaning ``at most as trusted as". 
Let $a$ range over processes, $\omega$ over objects, and $\labp,\labo$ over labels. 
Processes can fork new processes, create new objects, change the labels of processes and objects, and read, write, and execute objects. In particular, a process with label $\labp$ can:
\begin{compactenum2}
\item fork a new process $a(\labp)$; 
\item create a new object $\omega(\labp)$; 
\item lower its own label; 
\item change the label of an object $\omega(\labo)$ to $\labo'$ iff $\labo \sqcup \labo' \sqsubseteq \labp$; 
\item read an object $\omega(\labo)$;
\item write an object $\omega(\labo)$ iff $\labo \sqsubseteq \labp$;
\item execute an object $\omega(\labo)$ by lowering its own label to $\labp \sqcap \labo$.
\end{compactenum2} 
Rules (i) and (ii) are straightforward. 
Rule (iii) is guided by the principle of least
privilege~\cite{protection}, and is used in Windows Vista 
to implement a feature called \emph{user access control}
(UAC)~\cite{uac}. This feature lets users execute
commands with lower privileges when appropriate. For example, when a system administrator opens a new shell (typically with label 
$\mathsf{High}$), a new process is forked with label 
$\mathsf{Medium}$; the shell is then run by the new
process. When an Internet browser is opened, it is always run by a
new process whose label is lowered to $\mathsf{Low}$; thus any code
that gets run by the browser gets the label $\mathsf{Low}$---by Rule (i)---and any file that is downloaded by the browser gets the label $\mathsf{Low}$---by Rule~(ii).   

Rules (iv) and (v) are useful in various ways, but can be dangerous if not used carefully. (We show some attacks to illustrate this point below.) In particular, Rule (iv) allows unprotected objects to be protected by trusted processes by raising
their labels, and Rule~(v) allows processes to read objects at lower trust levels. At the same time, Rule (iv) facilitates dynamic access
control, and Rule (v) facilitates communication across trust boundaries. 

Rule (vi) protects objects from being written by processes at lower trust levels. Thus, for example, untrusted code
forked by a browser cannot affect local user files. User code cannot modify registry keys protected by a system administrator. 
Rule (vii) is part of UAC; it 
prevents users from accidentally launching less trusted executables with higher privileges. For example, a virus downloaded from the Internet cannot run
in a trusted user shell. Neither can system code dynamically link
user libraries.

\subsection{Some attacks}\label{vista}
We now show some attacks that remain possible in this environment. Basically, these attacks exploit Rules (iv) and (v) to bypass Rules (vi) and (vii). 
%
\begin{description}
\item[{\rm (}Write and copy{\rm )}] By Rule (vi), $a(\labp)$ cannot
  modify $\omega(\labo)$ if $\labp \sqsubset \labo$. However, $a(\labp)$ can modify 
  some object $\omega'(\labp)$, and then some process $b(\labo)$ can copy $\omega'(\labp)$'s content to $\omega(\labo)$. Thus, Rule (iv) can be exploited to bypass Rule (vi).
\item[{\rm (}Copy and execute{\rm )}] By Rule (vii), $a(\labp)$ cannot
  execute $\omega(\labo)$ at $\labp$ if $\labo \sqsubset \labp$. However, $a(\labp)$ can copy $\omega(\labo)$'s content to some object $\omega'(\labp)$ and then execute $\omega'(\labp)$. Thus, Rule (iv) can be exploited to bypass Rule (vii).
\item[{\rm (}Unprotect, write, and protect{\rm )}] By Rule (vi), $a(\labp)$ cannot
  modify $\omega(\labo)$ if $\labp \sqsubset \labo$. However, some process $b(\labo)$
  can unprotect $\omega(\labo)$ to $\omega(\labp)$, then $a(\labp)$ can modify 
   $\omega(\labp)$, and then $b(\labo)$ can protect $\omega(\labp)$ back to $\omega(\labo)$. Thus, Rule (v) can be exploited to bypass Rule (vi).
\item[{\rm (}Copy, protect, and execute{\rm )}] By Rule (vii), $a(\labp)$ cannot
  execute $\omega(\labo)$ at $\labp$ if $\labo \sqsubset \labp$. However, some process $b(\labo)$ can copy $\omega(\labo)$'s content to an object $\omega'(\labo)$, and then $a(\labp)$ can protect
  $\omega'(\labo)$ to $\omega'(\labp)$ and execute $\omega'(\labp)$. Thus, Rules (iv) and (v) can be exploited to bypass Rule (vii).
\end{description}
Next, we show that all of these attacks can violate DFI. At the same time, we observe that access
control forces the participation of a trusted process (one with the higher label) in
any such attack. 
\begin{itemize}
\item In ({\bf Write and copy}) or ({\bf Unprotect, write, and protect}), suppose that the contents of $\omega(\labo)$ are trusted, and $\labp$ is the label of untrusted code,  with $\labp \sqsubset \labo$. Then data can flow from $a(\labp)$ to $\omega(\labo)$, violating DFI, as above. Fortunately, some process $b(\labo)$ can be blamed here.
\item In ({\bf Copy and execute}) or ({\bf Copy, protect, and execute}), suppose that the contents of some object $\omega''(\labp)$ are trusted, and $\labo$ is the label of untrusted code, with $\labo \sqsubset \labp$. Then data can flow from some process $b(\labo)$  to $\omega''(\labp)$, violating DFI, as follows: $b(\labo)$ packs code to modify $\omega''(\labp)$ and writes the code to $\omega(\labo)$, and $a(\labp)$ unpacks and executes that code, as above. Fortunately, $a(\labp)$ can be blamed here.
\end{itemize}
Our type system can eliminate such attacks by restricting trusted processes (Section \ref{typ}).
(Obviously, the type system cannot restrict untrusted code running in the environment.) Conceptually, this guarantee can be cast as Wadler and Findler's ``\emph{well-typed programs can't be blamed}"~\cite{blame}. 
We rely on the fact that a trusted process can be blamed for any violation of DFI; it follows that if all trusted processes are well-typed, there cannot be any violation of DFI.

\section{A calculus for analyzing DFI on Windows Vista}\label{calc}
To formalize our approach, we now design a simple higher-order process calculus that simulates Windows Vista's security environment. We first introduce the syntax and informal semantics, and present some examples of programs and attacks in the language. We then present a formal semantics, guided by a precise characterization of explicit flows. 

\subsection{Syntax and informal semantics}

Several simplifications appear in the syntax of the language. We describe processes by their code. We use variables as object names, and let objects contain packed code or names of other objects. We enforce a mild syntactic restriction on nested packing, which makes typechecking significantly more efficient (Appendix \ref{algo}; also see below). 
Finally, we elide conditionals---for our purposes, the code 
$$\mathsf{if}~\mathtt{condition}~\mathsf{then}~a~\mathsf{else}~b$$
can be conservatively analyzed by composing $a$ and $b$ in parallel. (DFI is a \emph{safety property} in the sense of \cite{alpernschneider}, and 
the safety of the latter code implies that of the former. We discuss this point in more detail in Section \ref{introDFI}.) 

Values include variables, $\mathsf{unit}$, and packed expressions. Expressions include those for forking new processes, creating new objects, changing the labels of processes and objects, and reading, writing, and executing objects. 
%
They also include standard expressions for evaluation and returning
results (see Gordon and Hankin's concurrent object calculus \cite{gordon98concurrent}).  
\begin{defn}
\mycategory{f,g}{expression} \\
\entry{\fork f g}{fork} \\
\entry{t}{action} \\
\entry{\eval x f g}{evaluation} \\
\entry{r}{result} \\
\mycategory{t}{action} \\
\entry{\mathsf{new}(x\mbox{ \# }\labb)}{create object} \\
\entry{[\labp]~a}{change process label} \\
\entry{\langle\labo\rangle~\omega}{change object label}\\
\entry{!\omega}{read object} \\
\entry{\omega := x}{write object} \\
\entry{\mathsf{exec}~\omega}{execute object} \\
\mycategory{r}{result} \\
\entry{x,y,z,\dots,\omega}{variable} \\
\entry{\mathsf{unit}}{unit} \\
\\
\mycategory{a,b}{process} \\
\entry{\fork a b}{fork} \\
\entry{t}{action}  \\
\entry{\eval x a b}{evaluation} \\
\entry{u}{value} \\
\mycategory{u,v}{value} \\
\entry{r}{result} \\
\entry{\mathsf{pack}(f)}{packed expression} 
\end{defn}
Syntactically, we distinguish between processes and expressions: while every expression is a process, not every process is an expression. For example, the process $\mathsf{pack}(f)$ is not an expression, although the process $[\labp]~\mathsf{pack}(f)$ is. Expressions can be packed, but processes in general cannot. In particular, a process cannot be of the form $\mathsf{pack}(\mathsf{pack}(\dots))$. (Such a process can, however, be written as $\eval x {\mathsf{pack}(\dots)}\mathsf{pack}(x)$.) The benefits of this distinction become clear in Section \ref{Results}, where we discuss mechanical typechecking. However, for the bulk of the paper, the reader may ignore this distinction; indeed, neither the semantics nor the type system are affected by this distinction.

Processes have the following informal meanings. 
%
\begin{compact}
\item $\fork a b$ forks a new process $a$ with the current process label and continues as $b$ (see Rule (i)).
\item $\mathsf{new}(x\mbox{ \# }\labb)$ creates a new object $\omega$ with the current process label,  initializes $\omega$ with $x$, and returns $\omega$ (see Rule~(ii)); the annotation $\labb$ is used by the type system (Section \ref{typ}) and has no runtime significance. 
\item $[\labp]~a$ changes the current process label to $\labp$ and
  continues as
  $a$; it blocks if the current process label is lower than $\labp$ (see Rule~(iii)).
\item $\langle\labo\rangle~\omega$ changes $\omega$'s label to $\labo$ and returns $\mathsf{unit}$; it
  blocks if  $\omega$ is not bound to an object at runtime, or the current process label is lower than $\omega$'s label or $\labo$ (see Rule (iv)).
\item $!\omega$ returns the value stored in $\omega$; it blocks if
  $\omega$ is not bound to an object at runtime (see Rule (v)).
\item $\omega := x$ writes the value $x$ to $\omega$ and returns $\mathsf{unit}$; it blocks if
  $\omega$ is not bound to an object at runtime, or if the current process label is lower than $\omega$'s label (see Rule (vi)).
\item $\mathsf{exec}~\omega$ unpacks the value stored in $\omega$ to a
  process $f$, lowers the current process label with $\omega$'s label, and executes
  $f$; it blocks if $\omega$ is not bound to an object
 at runtime or if the value stored in $\omega$ is not a packed expression (see Rule (vii)).
\item $\eval x a b$ executes $a$, binds the value returned by $a$ to $x$, and
  continues as
  $b$ with $x$ bound.
\item $u$ returns itself.
\end{compact}

\subsection{Programming examples}

%
We now consider some programming examples in the language. We assume that $\mathsf{Low}$, $\mathsf{Medium}$, $\mathsf{High}$, and $\top$ are labels, ordered in the obvious way. We assume that the top-level process always runs with $\top$, which is the most trusted label. 

\begin{example}\label{eg:1} \rm Suppose that a $\mathsf{Medium}$ user opens an Internet browser $\mathtt{ie.exe}$ with $\mathsf{Low}$
privileges (recall UAC), and clicks on a $\mathtt{url}$ that contains $\mathtt{virus.exe}$; the virus contains code to overwrite
the command shell executable $\mathtt{cmd.exe}$, which has label $\top$. 
\begin{eqnarray*}
p_1\!\!\!\!&\triangleq&\!\!\!\eval{\mathtt{cmd.exe}}{\mathsf{new}(\dots\mbox{ \# }\top)}\\
\!\!\!\!&\!\!\!\!&\!\!\!\eval{\mathtt{url}}{[\mathsf{Low}]~\mathsf{new}(\dots\mbox{ \# }\mathsf{Low})}\\
\!\!\!\!&\!\!\!\!&\!\!\!\eval{\mathtt{binIE}}{\mathsf{pack}(\eval x {\:!\mathtt{url}}\mathsf{exec}~x)}\\
\!\!\!\!&\!\!\!\!&\!\!\!\eval{\mathtt{ie.exe}}{\mathsf{new}(\mathtt{binIE}\mbox{ \# }\top)}\\
\\
\!\!\!\!&\!\!\!\!&\!\!\![\mathsf{Medium}]~ (\dots ~\Rsh [\mathsf{Low}]~
\mathsf{exec}~\mathtt{ie.exe}) ~\Rsh\\
\!\!\!\!&\!\!\!\!&\!\!\![\mathsf{Low}]~ (\eval{\mathtt{binVirus}}{\mathsf{pack}(\mathtt{cmd.exe}
:= \dots)}\\
\!\!\!\!&\!\!\!\!&\!\!\!~~~~~~~~\:\:\:\:\eval{\mathtt{virus.exe}}{\mathsf{new}(\mathtt{binVirus}\mbox{ \# }\mathsf{Low})} \\
\!\!\!\!&\!\!\!\!&\!\!\!~~~~~~~~\:\:\:\:\mathtt{url} := 
\mathtt{virus.exe}~\Rsh\\
\!\!\!\!&\!\!\!\!&\!\!\!~~~~~~~~\:\:\:\:\dots) 
\end{eqnarray*}
This code may eventually reduce to 
\begin{eqnarray*}
q_1\!\!\!& \triangleq &\!\!\![\mathsf{Medium}]~ (\dots ~\Rsh [\mathsf{Low}]~
\mathtt{cmd.exe} := \dots)~\Rsh \\
&&\!\!\![\mathsf{Low}]~ (\dots)
\end{eqnarray*}
However, at this point the write to $\mathtt{cmd.exe}$ blocks due to
access control. (Recall that a process with label $\mathsf{Low}$ cannot write to an
object with label $\top$.)
\end{example}

\begin{example}\label{eg:2} \rm
Next, consider the following attack, based on the ({\bf Copy, protect, and execute}) attack in Section \ref{vista}. 
A $\mathsf{Medium}$ user 
downloads a virus from the Internet that contains code to erase the user's home directory ($\mathtt{home}$), and saves it by default in $\mathtt{setup.exe}$. A $\mathsf{High}$
administrator protects and executes $\mathtt{setup.exe}$.
\begin{eqnarray*}
p_2\!\!\!\!&\triangleq&\!\!\!\eval{\mathtt{url}}{[\mathsf{Low}]~\mathsf{new}(\dots\mbox{ \# }\mathsf{Low})}\\
\!\!\!\!&\!\!\!\!&\!\!\!\eval{\mathtt{setup.exe}}{[\mathsf{Low}]~\mathsf{new}(\dots\mbox{ \# }\mathsf{Low})}\\
\!\!\!\!&\!\!\!\!&\!\!\!\eval{\mathtt{binIE}}{\mathsf{pack}(\eval z {~!\mathtt{url}}\\
\!\!\!\!&\!\!\!\!&\!\!\!~~~~~~~~~~~~~~~~~~~~~~~~~\:\:\:\:\:\:\eval x{~!z}\mathtt{setup.exe}:=~x)}\\
\!\!\!\!&\!\!\!\!&\!\!\!\eval{\mathtt{ie.exe}}{\mathsf{new}(\mathtt{binIE}\mbox{ \# }\top)}\\
\!\!\!\!&\!\!\!\!&\!\!\!\eval{\mathtt{home}}{[\mathsf{Medium}]~\mathsf{new}(\dots\mbox{ \# }\mathsf{Medium})}\\
\!\!\!\!&\!\!\!\!&\!\!\!\eval{\mathtt{empty}}{\mathsf{unit}}\\
\\
\!\!\!\!&\!\!\!\! &\!\!\![\mathsf{High}]~ (\dots \Rsh \\
\!\!\!\!&\!\!\!\! &\!\!\!~~~~~~~~~~~~\:\eval \_ {\langle\mathsf{High}\rangle~\mathtt{setup.exe}}\\
\!\!\!\!&\!\!\!\! &\!\!\!~~~~~~~~~~~~\:\mathsf{exec}~\mathtt{setup.exe})~\Rsh \\
\!\!\!\!&\!\!\!\! &\!\!\![\mathsf{Medium}]~ (\dots ~\Rsh~ [\mathsf{Low}]~
\mathsf{exec}~\mathtt{ie.exe}) ~\Rsh\\
\!\!\!\!&\!\!\!\! &\!\!\![\mathsf{Low}]~ (\eval{\mathtt{binVirus}}{\mathsf{pack}(\mathtt{home}
:= \mathtt{empty})}\\
\!\!\!\!&\!\!\!\! &\!\!\!~~~~~~~~\:\:\:\:\eval{\mathtt{virus.exe}}{\mathsf{new}(\mathtt{binVirus}\mbox{ \# }\mathsf{Low})} \\
\!\!\!\!&\!\!\!\!&\!\!\!~~~~~~~~\:\:\:\:\mathtt{url} := 
\mathtt{virus.exe}~\Rsh \\
\!\!\!\!&\!\!\!\!&\!\!\!~~~~~~~~\:\:\:\:\dots)
\end{eqnarray*}
This code may eventually reduce to 
\begin{eqnarray*}
q_2 \!\!\!& \triangleq &\!\!\![\mathsf{High}]~ (\dots~\Rsh~\mathtt{home}
:= \mathtt{empty})~\Rsh \\
&&\!\!\![\mathsf{Medium}]~ (\dots)~\Rsh \\
&&\!\!\![\mathsf{Low}]~ (\dots)
\end{eqnarray*}
The user's home directory may be erased at this point.  (Recall that access control does not prevent a process with label $\mathsf{High}$ from writing to an
object with label $\mathsf{Medium}$.)
\end{example}

\subsection{An overview of DFI}\label{introDFI}
Informally, DFI requires that objects whose contents are trusted at some label $\labb$ never contain values that flow from labels lower than $\labb$. In Example~\ref{eg:1}, we trust the contents of $\mathtt{cmd.exe}$ at label $\top$, as declared by the static annotation $\top$. DFI is \emph{not} violated in this example, since access control prevents the flow of data from $\mathsf{Low}$ to $\mathtt{cmd.exe}$. On the other hand, in Example \ref{eg:2}, we trust the contents of $\mathtt{home}$ at label $\mathsf{Medium}$. DFI \emph{is} violated in this example, since the value $\mathtt{empty}$ flows from $\mathsf{Low}$ to $\mathtt{home}$. 

By design, DFI is a safety property  \cite{alpernschneider}---roughly, it can be defined as a set of behaviors such that for any behavior that not in that set, there is some finite prefix of that behavior that is not in that set. To that end, DFI considers only \emph{explicit} flows of data. Denning and Denning characterizes explicit flows~\cite{denningcert} roughly as follows: a flow of $x$ is explicit if and only if the flow depends abstractly on $x$ (that is, it depends on the existence of $x$, but not on the value $x$). Thus, for example, the violation of DFI in Example \ref{eg:2} does not depend on the value $\mathtt{empty}$---\emph{any} other value causes the same violation. Conversely, $\mathtt{empty}$ is not dangerous in itself. Consider the reduced process $q_2$
in Example \ref{eg:2}. Without any knowledge of execution history, we cannot conclude that DFI is violated in $q_2$. Indeed, it is perfectly legitimate for a $\mathsf{High}$-process to execute the code $$\mathtt{home} := \mathtt{empty}$$ 
intentionally, say as part of administration. However, in Example~\ref{eg:2}, we know that this code is executed by unpacking some code designed by a $\mathsf{Low}$-process. The violation of DFI is \emph{due to this history}. 

It follows that in order to detect violations of DFI, we must distinguish between various instances of a value, and track the sources of those instances during execution. 
We maintain this execution history in
the operational semantics (Section \ref{semantics}), by a technique based on explicit substitution \cite{abadi90explicit}. 

Before we move on, let us ease the tension between DFI and conditionals. In general, conditionals can cause implicit flows~\cite{denningcert}; a flow of $x$ can depend on the value $x$ if $x$ appears in the condition of some code that causes that flow. For example, the code
$$\mathsf{if}~x = \mathtt{zero}~\mathsf{then}~\omega := \mathtt{zero}~\mathsf{else}~\omega := \mathtt{one}$$
causes an implicit flow of $x$ to $\omega$ that depends on the value $x$. We can abstract away this dependency by interpreting the code $\mathsf{if}~\mathtt{condition}~\mathsf{then}~a~\mathsf{else}~b$ as the code $\fork a b$. Recall that DFI is a safety property. Following \cite{lamport77}, the safety of $\fork a b$ can be expressed by the logical formula $F \triangleq F_a \wedge F_b$, where $F_a$ is the formula that expresses the safety of $a$, and $F_b$ is the formula that expresses the safety of $b$. Likewise, the safety of $\mathsf{if}~\mathtt{condition}~\mathsf{then}~a~\mathsf{else}~b$ can be expressed by the formula $F' \triangleq (\mathtt{condition} \Rightarrow F_a) \wedge (\neg \mathtt{condition} \Rightarrow F_b)$. Clearly, we have  $F \Rightarrow F'$, so that the code $\mathsf{if}~\mathtt{condition}~\mathsf{then}~a~\mathsf{else}~b$ is a refinement of the code $\fork a b$. It is well-known that safety is preserved under refinement \cite{lamport77}.

But implicit flows are of serious concern in many applications; one may wonder whether focusing on explicit flows is even desirable. Consider the code above; the implicit flow from $x$ to $\omega$ violates noninterference, if $x$ is an untrusted value and the contents of $\omega$ are trusted. In contrast, DFI is \emph{not} violated in the interpreted code 
$$\omega := \mathtt{zero}~\Rsh~\omega := \mathtt{one}$$
if $\mathtt{zero}$ and $\mathtt{one}$ are trusted values. Clearly, DFI ignores the implicit flow from $x$ to $\omega$. But this may be fine---DFI can be used to prove an invariant such as ``the contents of $\omega$ are always boolean values". Note that the code
$$\omega := x$$
does not maintain this invariant, since $x$ may be an arbitrary value. Thankfully, DFI \emph{is} violated in this code.

\subsection{An operational semantics that tracks explicit flows}\label{semantics}
\begin{figure}
\cenvvv{Local reduction}{a \actsub{\labp} b\textcolor{white}{p}}{
({\bf Reduct evaluate})\vspace{-1mm}
$$\infer
    {}
    {\eval x u a \actsub{\labp} \new{x/u@\labp} a}
$$
~

({\bf Reduct new})\vspace{-1mm}
$$\infer
    {}
    {\mathsf{new}(x\mbox{ \# }\labb) \actsub{\labp} \new{\omega/\mathsf{new}(x\mbox{ \# }\labb) @\labp}(\omega \store{\labp} x \Rsh \omega)}
$$
~

({\bf Reduct read})\vspace{-2.5mm}
$$
\infer
    {\omega \stackrel\sigma= \omega'}
    {\fork{\omega \store{\labo} x} ~!\omega'  \actsub{\labp} \fork{\omega \store{\labo} x}x}
$$
~

({\bf Reduct write})\vspace{-1mm}
$$\infer
    {\omega \stackrel\sigma= \omega' \\ \labo \sqsubseteq \labp}
    {\fork{\omega \store{\labo} \_~} \omega' := x \actsub{\labp} \fork{\omega \store{\labo} x} \mathsf{unit}}
$$
~

({\bf Reduct execute})\vspace{-1mm}
$$\infer
    {\omega \stackrel\sigma= \omega' \\ \mathsf{pack}(f) \in \sigma(x) \\ \labp' = \labp \sqcap \labo}
    {\fork{\omega \store{\labo} x}\mathsf{exec}~\omega' \actsub{\labp} \fork{\omega \store{\labo} x}[\labp']~f}
$$
~

({\bf Reduct un/protect})\vspace{-1mm}
$$\infer
    {\omega \stackrel\sigma= \omega' \\ \labo \sqcup \labo' \sqsubseteq \labp}
    {\fork{\omega \store{\labo} x}\langle\labo'\rangle\: \omega' \actsub{\labp} \fork{\omega \store{\labo'} x}\mathsf{unit}}
$$
}
\vspace{-2mm}
\cenvvv{Structural equivalence}{a \equiv b}{
%
({\bf Struct bind})\vspace{-1mm}
$$\infer
    {}
    { \sctxr {a\{x/y\}} {\labp',\sigma'}  \equiv \sctxr{\new {x/y@\labp'} a} {\labp',\sigma'} }
$$
~

({\bf Struct substitution})\vspace{-0mm}
$$\infer
    {x \notin \fv(\mathcal E_{\labp,\sigma}) \cup \bv(\mathcal E_{\labp,\sigma}) \\ \fv(\mu) \cap \bv(\mathcal E_{\labp,\sigma}) = \varnothing}
    { \sctxr{\new {x/\mu@\labp''} a} {\labp',\sigma'} \equiv \new {x/\mu@\labp''} \llctx{\labp,\{x/\mu@\labp''\}\cup \sigma} a {\labp',\sigma'} }
$$
~

({\bf Struct fork})\vspace{-1mm}
$$\infer
    {\fv(a) \cap \bv(\mathcal E_{\labp,\sigma}) = \varnothing}
    {\sctxr{\fork a b} {\labp,\sigma'} \equiv \fork a \sctxr b {\labp,\sigma'}}
$$
~

({\bf Struct store})\vspace{-1mm}
$$\infer
    {}
    { [\labp]~(\fork{\omega \store{\labo} x} a) \equiv \fork{\omega \store{\labo} x}[\labp]~a}
$$
~

({\bf Struct equiv})\vspace{-2.5mm}
$$\infer
    {}
    {\equiv\mbox{ is an equivalence}}
$$
}
\vspace{-2mm}
\cenvvv{Global reduction}{a \actsub{\labp} b\textcolor{white}{p}}{
({\bf Reduct context})\vspace{-1mm}
$$
\infer
    {a \action{\labp';\sigma'} b}
    {\sctxr a {\labp';\sigma'}\action{\labp;\sigma} \sctxr b {\labp';\sigma'}}
$$
~

({\bf Reduct congruence})\vspace{-1mm}
$$\infer
    {a \equiv a' \\ a' \actsub{\labp} b' \\ b' \equiv b}
    {a \actsub{\labp} b}
$$
}
\end{figure}
%
We now present a chemical-style operational semantics for the language, that tracks explicit flows.\footnote{This presentation is particularly convenient for defining and proving DFI; of course, a concrete implementation of the language may rely on a lighter semantics that does not track explicit flows.}
We begin by extending the syntax with some auxiliary forms.
\begin{defn}
\mycategory{a,b}{process} \\
\entry{\cdots}{source process}\\
\entry{\omega \store{\labo} x}{store} \\
\entry{\new {x/\mu@{\labp}} a}{explicit substitution}  \\
\mycategory{\mu}{substituted value} \\
\entry{u}{value}\\
\entry{\mathsf{new}(x\mbox{ \# }\labb)}{object initialization}
\end{defn} 
The process $\omega \store{\labo} x$ asserts that the object $\omega$ contains $x$ and is protected with label $\labo$. A key feature of the semantics is that objects store values ``by instance"---only variables may appear in stores. 
We use explicit substitution to track and distinguish
between the sources of various instances of a substituted value. 
Specifically, the process $\new{x/\mu@\labp}a$ creates a fresh variable $x$, records that $x$ is bound to $\mu$ by a
process with label $\labp$, and continues as $a$ with $x$ bound. Here $x$ is an \emph{instance} of $\mu$ and $\labp$ is the \emph{source} of $x$. If $\mu$ is a value, then this process is behaviorally equivalent to $a$ with $x$ substituted by $\mu$. 
For example, in Example \ref{eg:2} the source of the instance of $\mathtt{empty}$ in $\mathtt{binVirus}$ is $\mathsf{Low}$; this fact is described by rewriting the process $q_2$ as 
$$\new {x/\mathtt{empty}@\mathsf{Low}} [\mathsf{High}]~(\dots \Rsh~ \mathtt{home}
:= x)  \Rsh \dots$$
DFI prevents this particular instance ($x$) of $\mathtt{empty}$ from being written to $\mathtt{home}$; but it allows other instances whose sources are at least as trusted as $\mathsf{Medium}$. The rewriting follows a structural equivalence rule (\srule{bind}), explained later in the section.

While explicit substitution has been previously used in language implementations, we seem to be the first to adapt this device to track data flow in a concurrent language. In particular, we use explicit substitution both to specify DFI (in Definitions \ref{flowsdef} and \ref{locintdef1}) and to verify it statically (in proofs of Theorems \ref{subjred} and \ref{mainthm}).
We defer a more detailed discussion on this technique to Section~\ref{concl}.

We call sets of the form $\{x_1/\mu_1@\labp_1,\dots,x_k/\mu_k@\labp_k\}$
\emph{substitution environments}. 
\begin{definition}[Explicit flows]\label{flowsdef} A variable $x$ flows from a label $\labp$ or lower in a substitution environment $\sigma$, written $x \stackrel\sigma\blacktriangledown \labp$, if $x/\mu@\labp' \in \sigma$ for some $\mu$ and $\labp'$ such that either $\labp' \sqsubseteq \labp$, or $\mu$ is a variable and (inductively) $\mu \stackrel\sigma\blacktriangledown  \labp$.
\end{definition}
%
In other words, $x$ flows from a label $\labp$ or lower if $x$ is an instance of a value substituted at $\labp$ or lower. 
In Definition~\ref{locintdef1} below, we formalize DFI as a property of objects, as follows: \emph{an object is protected from label $\lab$ if it never contains instances that flow from $\lab$ or lower.} 
We define $\sigma(x)$ to be the set of values in $\sigma$ that $x$ is an instance of: $x \in \sigma(x)$, and if (inductively) $y \in \sigma(x)$ and $y/u@~\_ \in \sigma$ for some $y$ and $u$, then $u \in \sigma(x)$. 
The operational semantics ensures that substitution environments accurately associate instances of values with their runtime sources. 

We now present rules for local reduction, structural equivalence, and global reduction.
Reductions are of the form $a
\action{\labp;\sigma} b$, meaning that ``process $a$ 
may reduce to process $b$ with label $\labp$ in
substitution environment $\sigma$''. Structural equivalences are of the form $a \equiv b$, meaning that ``process $a$ may be rewritten as process $b$''.
The notions
of free and bound variables ($\fv$ and $\bv$) are
standard. 
We write $x \stackrel\sigma= y$ if $\sigma(x) \cap \sigma(y) \neq \varnothing$, that is, there is a value that both $x$ and $y$ are instances of.

We first look at the local reduction rules.
In (\rrule{evaluate\-}), a substitution binds $x$ to the intermediate value $u$ and associates $x$ with its
runtime source $\labp$. 
(\rrule{new}) creates a new store denoted by a fresh variable $\omega$, initializes the store, and returns $\omega$; a substitution binds $\omega$ to the initialization of the new object and associates $\omega$ with its runtime source $\labp$. The value $x$ and the trust annotation $\labb$ in the initialization are used by the type system (Section \ref{typ}).
The remaining local reduction rules describe reactions with a store, following the informal semantics.

Next, we define evaluation contexts \cite{evalcontext}. An evaluation context is of the form $\mathcal E_{\labp;\sigma}$, and contains a hole of the form $\bullet_{\labp';\sigma'}$; the context yields a process that executes with label
$\labp$ in substitution environment $\sigma$, if the hole is plugged by a process that executes with label $\labp'$ in substitution environment $\sigma'$.
\begin{defn2}
\mycategory{\mathcal E_{\labp;\sigma}}{evaluation context} \\
\entry{\bullet_{\labp;\sigma}}{hole} \\
\entry{\eval x {\mathcal E_{\labp;\sigma}} b}{sequential evaluation} \\
\entry{\fork{\mathcal E_{\labp;\sigma}} b}{fork left} \\
\entry{\fork a \mathcal E_{\labp;\sigma}}{fork right} \\
\entry{\new {x/\mu @\labp'} \mathcal E_{\labp;\{x/\mu@\labp'\}\cup \sigma}}{explicit substitution} \\
\entry{[\labp']~\mathcal E_{\labp';\sigma}~~~~~(\labp' \sqsubseteq \labp)}{lowering of process label}
\end{defn2}
%
Evaluation can proceed sequentially inside $\mathsf{let}$ processes, and in parallel under forks \cite{gordon98concurrent}; it can also proceed under
explicit substitutions and lowering of process labels. 
In particular, note how evaluation contexts build substitution environments from explicit substitutions, and labels from changes of process labels. 
We denote by $\sctxr a {\labp';\sigma'}$ the process obtained by plugging the
hole $\bullet_{\labp';\sigma'}$ in $\mathcal E_{\labp;\sigma}$ with $a$. 

Next, we look at the structural
equivalence and global reduction rules. 
%
%
In (\srule{bind}), $a\{x/y\}$ is the process obtained from $a$ by the usual capture-avoiding substitution of $x$ by $y$. 
The rule states that explicit substitution may \emph{invert} usual substitution to create instances as required. In particular, variables that appear in packed code can be associated with the label of the process that packs that code, even though those variables may be bound later---by (\rrule{evaluate})---when that code is eventually unpacked at some other label. For example, the instance of $\mathtt{empty}$ in $\mathtt{binVirus}$ may be correctly associated with $\mathsf{Low}$ (the label at which it is packed) instead of $\mathsf{High}$ (the label at which it is unpacked). Thus, in combination, the rules (\rrule{evaluate}) and (\srule{bind}) track precise sources of values by explicit substitution. 

By (\srule{substitution}), substitutions can float across contexts under standard scoping restrictions. 
By (\srule{fork}), forked processes can float across contexts \cite{gordon98concurrent}, but must remain under the same
process label. By (\srule{store}), stores can be shared across further contexts. 

Reduction is extended with contexts and structural equivalence in the natural way.

Finally, we formalize DFI in our language, as promised. 
\begin{definition}[DFI]\label{locintdef1} The object $\omega$ is protected from label $\lab$ by process $a$
  if there is no process $b$, substitution environment $\sigma$, and instance $x$ such that $a \Rsh [\lab]~b
  \action{\top,\varnothing}\!\!\!\!{}^\star ~\llctx{\top,\varnothing}{\omega\store{\_} x}{\top,\sigma}$ and $x\stackrel\sigma\blacktriangledown \lab$.
\end{definition}
%

\section{A type system to enforce DFI}\label{typ}
We now show a type system to enforce DFI in the language. (The formal protection guarantee for well-typed code appears in Section~\ref{Results}.) We begin by introducing types and typing
judgments. We then present typing rules and informally
explain their properties. Finally, we consider some examples of typechecking. 
An efficient algorithm for typechecking is outlined in Appendix \ref{algo}. 

\subsection{Types and effects}
\begin{figure}
\cenvvv{Core typing judgments}{\Gamma \vdash_{\labp} a : T}{
({\bf Typ unit})\vspace{-1mm}
$$\infer
{}{\Gamma \vdash_{\labp} \mathsf{unit} : \mathbf{Unit}^{\labp}}
$$
~

({\bf Typ variable})\vspace{-1mm}
$$\infer
    {x : \tau^{\labt} \in \Gamma}
    {\Gamma \vdash_{\labp} x : \tau^{\labt \sqcap \labp}}
$$
~

({\bf Typ fork})\vspace{-0mm}
$$\infer
    {\Gamma \vdash_{\labp} a : \_ \\
    \Gamma \vdash_{\labp} b : T}
    {\Gamma \vdash_{\labp} \fork a b : T}
$$
~

({\bf Typ limit})\vspace{-1mm}
$$\infer
    {
    \Gamma
    \vdash_{\labp'} a : T}
    {\Gamma \vdash_{\labp} [\labp']~a : T}
$$
~

({\bf Typ evaluate})\vspace{1mm}
$$
\infer
    {\Gamma \vdash_{\labp} a : T' \\
    \Gamma,x: T' \vdash_{\labp} b : T}
    {\Gamma \vdash_{\labp} \eval x a b : T}
$$
~

({\bf Typ substitute})\vspace{1mm}
$$\infer
    {\Gamma \vdash_{\labp'} \mu : T' \\
    ~~~\Gamma, x: T' \vdash_{\labp} a : T}
    {\Gamma \vdash_{\labp} \new{x/\mu @\labp'} a : T}
$$
~

({\bf Typ store})\vspace{1mm}
$$\infer
    {\{\omega : \mathbf{Obj}(\tau^{\labb})^\_, x : \tau^{\labt}\} \subseteq \Gamma \\
    \labb \sqsubseteq \labo \sqcap \labt}
    {\Gamma \vdash_{\labp} \omega \store{\labo} x : \_^{\labp}}
$$
~

({\bf Typ new})\vspace{-1mm}
$$\infer
    {\Gamma \vdash_{\labp} x : \tau^{\labt} \\ 
    \labb \sqsubseteq \labt}
    {\Gamma \vdash_{\labp} \mathsf{new}(x\mbox{ \# }\labb) : \mathbf{Obj}(\tau^{\labb})^{\labp}}
$$
~

({\bf Typ pack})\vspace{-0mm}
$$\infer
    {\Gamma \vdash_{\labp'} f : T\\ 
    \Box f}
    {\Gamma \vdash_{\labp} \mathsf{pack}(f) : \nabla_{\labp'}.~\mathbf{Bin}(T)^{\labp}}
$$
%
~

({\bf Typ un/protect})\vspace{1mm}
$$
\inferrule*[right= ~\rem{\ast\labp\Rightarrow\ast\labt} ]
    {\Gamma \vdash_{\labp} \omega :
    \mathbf{Obj}(\_^{\labb})^{\labt} \\
    \labb \sqsubseteq \labo
    }
    {\Gamma \vdash_{\labp} \langle\labo\rangle~\omega : \mathbf{Unit}^\labp}
$$
~

({\bf Typ write})\vspace{1mm}
$$\inferrule*[right= ~\rem{\ast\labp \Rightarrow \ast\labt}]
    {\Gamma \vdash_{\labp} \omega :
    \mathbf{Obj}(\tau^{\labb})^{\labt} \\
    \Gamma \vdash_{\labp} x : \tau^{\labt'} \\
     \labb \sqsubseteq \labt' 
    }
    {\Gamma \vdash_{\labp} \omega := x : \mathbf{Unit}^\labp}
$$
~

({\bf Typ read})\vspace{1mm}
$$\inferrule*[right= ~\rem{\ast(\labp \sqcap \labb)\Rightarrow \ast\labt}]
    {\omega : \mathbf{Obj}(\tau^{\labb})^{\labt} \in \Gamma
    }
    {\Gamma \vdash_{\labp} ~!\omega : \tau^{\labb \sqcap \labp}}
$$
~

({\bf Typ execute})\vspace{1mm}
$$\inferrule*[right= ~\rem{\ast\labp \Rightarrow \ast \labt}]
    {\omega : \mathbf{Obj}((\nabla_{\labp'}.~\mathbf{Bin}(\tau^{\labt'}))^{~\labb})^{\labt} \in \Gamma \quad~~~
    \labp \sqsubseteq \labp' \sqcap \labb
     }
    {\Gamma \vdash_{\labp} \mathsf{exec}~\omega : \tau^{\labt' \sqcap \labp}}
$$
}
\end{figure}
The core grammar of types is shown below. Here effects are simply labels; these labels belong to the same ordering $\sqsubseteq$ as in the operational semantics. 
\begin{defn}
\mycategory{\tau}{type}\\
\entry{\mathbf{Obj}(T)}{object} \\
\entry{\nabla_{\labp}.~\mathbf{Bin}(T)}{packed code} \\
\entry{\mathbf{Unit}}{unit} \\
\mycategory{T}{static approximation}\\
\entry{\tau^\labt}{type and effect}
\end{defn}
\begin{compact}
\item The type $\mathbf{Obj}(\tau^{\labb})$ is given to an object that
  contains values of type $\tau$. Such contents may not flow from labels lower than $\labb$; in other words, $\labb$ indicates the trust on the contents of this object. DFI follows from the soundness of object types.
\item The type $\nabla_{\labp}.~\mathbf{Bin}(\tau^{\labt})$ is given to packed
  code that can be run with label $\labp$. Values returned by the code must be of type $\tau$ and
  may not flow from labels lower than~$\labt$. In fact, our type system admits a subtyping rule that allows such code to be run in a typesafe manner with any label that is at most $\labp$.
\item The effect $\labt$ is given to a value that does not flow from labels lower than $\labt$. 
\end{compact}
%
When creating an object, the programmer declares the trust on the contents of that object. 
Roughly, an object returned by $\mathsf{new}(\_\mbox{ \# }\labb)$ gets a type $\mathbf{Obj}(\_^{\labb})$.
For example, in Examples \ref{eg:1} and \ref{eg:2}, we declare the trust $\top$ on the contents of $\mathtt{cmd.exe}$ and the trust $\mathsf{Medium}$ on the contents of $\mathtt{home}$.

A typing environment $\Gamma$ contains 
typing hypotheses of the form $x : T$. We assume that any variable has at most one typing hypothesis in $\Gamma$, and define $\mathsf{dom}(\Gamma)$ as the set of variables that have
typing hypotheses in $\Gamma$. 
A typing judgment is of the
form $\Gamma \vdash_{\labp} a : T$, where $\labp$ is the label of the process $a$, $T$ is the type and effect of values returned by $a$, and  $\fv(a) \subseteq \dom(\Gamma)$.
%

\subsection{Core typing rules}\label{core}
%


In the previous page, we present typing rules that enforce the core static discipline required for our protection guarantee. Some of these rules have side conditions that involve a predicate $\ast$ on labels. These conditions, which are marked in \textshade[.91]{sharpcorners}{\gdef\outlineboxwidth{0.01}shaded boxes}, are ignored in our first reading of these rules. (The predicate $\ast$ is true everywhere in the absence of a special label $\bot$, introduced later in the section.) One of the rules has a condition that involves a predicate $\Box$ on expressions; we introduce that predicate in the discussion below.
%
The typing rules preserve several invariants.
\begin{compactenum3}
\item Code that runs with a label $\labp$ cannot return values that have effects higher than $\labp$.
\item The contents of an object of type $\mathbf{Obj}(\_^{\labb})$ cannot have effects lower than $\labb$. 
\item The dynamic label that protects an object of type $\mathbf{Obj}(\_^{\labb})$ cannot be lower than $\labb$. 
\item An object of type $\mathbf{Obj}(\_^{\labb})$ cannot be created at a label lower than $\labb$. 
\item Packed code of type $\nabla_{\labp}.~\mathbf{Bin}(\_)$ must remain well-typed when unpacked at any label lower than $\labp$.
\end{compactenum3}
%
Invariant (1) follows from our interpretation of effects. To preserve this invariant in \trule{variable}, for example, the effect of $x$ at $\labp$ is obtained by lowering $x$'s effect in the typing environment with $\labp$.

In \trule{store}, typechecking is independent of the process label, that is, a store is well-typed if and only if it is so at any process label; recall that by (\srule{store}) stores can float across contexts, and typing must be preserved by structural equivalence. Further, \trule{store} introduces Invariants (2) and (3). Invariant (2) follows from our interpretation of static trust annotations. To preserve this invariant we require Invariant (3), which ensures that access control prevents code running with labels less trusted than $\labb$ from writing to objects whose contents are trusted at $\labb$. 

By \trule{new}, the effect $\labt$ of the initial content of a new object cannot be lower than $\labb$. Recall that by (\rrule{new}), the new object is protected with the process label $\labp$; since $\labp \sqsupseteq \labt$ by Invariant (1), we have $\labp \sqsupseteq \labb$, so that both Invariants (2) and (3) are preserved. Conversely, if $\labp \sqsubset \labb$ then the process does not typecheck; Invariant (4) follows. 

Let us now look carefully at the other rules relevant to Invariants (2) and (3); these rules---combined with access control---are the crux of enforcing DFI. 
\trule{write} preserves Invariant (2), restricting trusted code from writing values to $\omega$ that may flow from labels lower than $\labb$. (Such code may not be restricted by access control.) 
Conversely, access control prevents code with labels lower than $\labb$ from writing to $\omega$, since by Invariant (3), $\omega$'s label is at least as trusted as~$\labb$. 
\trule{un/protect} preserves Invariant (3), allowing $\omega$'s label to be either raised or lowered without falling below~$\labb$. 
In \trule{read}, the effect of a value read from $\omega$ at~$\labp$ is approximated by $\labb$---the least trusted
label from which $\omega$'s contents may flow---and further lowered with $\labp$ to preserve Invariant~(1). 

In \trule{pack}, packing code requires work akin to proof-carrying
code~\cite{pcc}. Type safety for the code is proved and ``carried" in its type
$\nabla_{\labp'}.~\mathbf{Bin}(T)$, independently of the current process label. Specifically, it is proved that when the packed code
is unpacked by a process with label $\labp'$, the value of executing that code has type and effect $T$. In Section \ref{Results} we show that such a proof in fact allows the packed code
to be unpacked by any process with label $\labp \sqsubseteq \labp'$, and the type and effect of the value of executing that code can be related to $T$ (Invariant (5)). 
This invariant is key to decidable and efficient typechecking (Appendix \ref{algo}). Of course, code may be packed to run only at specific process labels, by requiring the appropriate label changes.

Preserving Invariant (5) entails, in particular, preserving Invariant (4) at all labels $\labp \sqsubseteq \labp'$. Since a $\mathsf{new}$ expression that is not guarded by a change of the process label may be run with any label $\labp$, that expression must place the least possible trust on the contents of the object it creates. 
This condition is enforced  by predicate  $\Box$:
%
\begin{eqnarray*}
\Box \mathsf{new}(x\mbox{ \# }\labb) & \triangleq & \forall \labp.~\labb \sqsubseteq \labp \\ 
\Box(\fork f g) &\triangleq & \Box f  \wedge \Box g \\ 
\Box(\eval x f g)  &\triangleq &  \Box f  \wedge \Box g \\ 
\Box(\dots) &\triangleq &  \mathtt{true} 
\end{eqnarray*}
%
\trule{execute} relies on Invariant (5); further, it checks that the label at which the code is unpacked ($\labp$) is at most as trusted as the label at which the code may have been packed (approximated by $\labb$). This check prevents privilege escalation---code that would perhaps block if run with a lower label cannot be packed to run with a higher label. For example, recall that in Example \ref{eg:2}, the code $\mathtt{binVirus}$
is packed at $\mathsf{Low}$ and then copied into $\mathtt{setup.exe}$. 
While a $\mathsf{High}$-process can legitimately execute $\mathtt{home} := \mathtt{empty}$ (so that the code is typed and is not blocked by access control), it should not run that code by unpacking $\mathtt{binVirus}$ from $\mathtt{setup.exe}$. 
The type system prevents this violation. Let $\mathtt{setup.exe}$ be of type $\mathbf{Obj}((\nabla_{\_}.~\mathbf{Bin}(\_))^{\labb})$. Then \trule{store} requires that $\labb \sqsubseteq \mathsf{Low}$, and \trule{execute} requires that $\mathsf{High} \sqsubseteq \labb$ (contradiction). 

Because we do not maintain an upper bound on the dynamic label of an executable, we cannot rely on the lowering of the process label in (\rrule{execute}) to prevent privilege escalation. (While it is possible to extend our type system to maintain such upper bounds, such an extension does not let us typecheck any more correct programs than we already do.) In Section \ref{Results}, we show that the lowering of the process label can in fact be safely
eliminated. 

In \trule{evaluate}, typing proceeds sequentially, propagating the type and effect of the intermediate process to the continuation.
\trule{substitution} is similar, except that the substituted value is typed under the process label recorded in the
substitution, rather than under the current process label. In \trule{limit}, the continuation is typed under the
changed process label. In \trule{fork}, the forked process is typed under the current process label.

\subsection{Typing rules for stuck code}\label{stuck}

\begin{figure}
\cenvvv{Stuck typing judgments}{\Gamma \vdash_{\labp} a : \mathbf{Stuck}}{
%
({\bf Typ escalate stuck})\vspace{-2mm}
$$\infer
    {\labp \sqsubset \labp'}
    {\Gamma \vdash_{\labp} [\labp']~a : \mathbf{Stuck}}
$$
~

({\bf Typ write stuck})\vspace{-1mm}
$$\inferrule*[right= ~\rem{\ast\labt}]
    {\omega : \mathbf{Obj}(\_^{\labb})^{\labt} \in \Gamma \\
    \labp \sqsubset \labb 
     }
    {\Gamma \vdash_{\labp} \omega := x : \mathbf{Stuck}}
$$
~

({\bf Typ un/protect stuck})\vspace{0mm}
$$
\inferrule*[right= ~\rem{\ast \labt}]
    {\omega : \mathbf{Obj}(\_^{\labb})^{\labt} \in \Gamma \\
    \labp \sqsubset \labb \sqcup \labo 
     }
    {\Gamma \vdash_{\labp} \langle\labo\rangle~\omega : \mathbf{Stuck}}
$$
~

({\bf Typ subsumption stuck-I})\vspace{-1mm}
$$
\infer
	{\_ : \mathbf{Stuck} \in \Gamma}
    {\Gamma \vdash_{\labp} a : \mathbf{Stuck}}
$$
~

({\bf Typ subsumption stuck-II})\vspace{0mm}
$$\infer
	{\Gamma \vdash_{\labp} a : \mathbf{Stuck}}
    {\Gamma \vdash_{\labp} a : T}
$$
}
\end{figure}

While the rules above rely on access control for soundness, they do not \emph{exploit} runtime protection provided by access control to typecheck more programs. For example, the reduced process $q_1$
in Example~\ref{eg:1} cannot yet be typed, although we have checked that DFI is not violated in $q_1$. Below we introduce \emph{stuck typing} to identify processes that provably block by access control at runtime. Stuck typing allows us to soundly type more programs by composition. 
(The general principle that is followed here is that narrowing the set of possible execution paths improves the precision of the analysis.) 
This powerful technique of combining static typing and dynamic access control for runtime protection is quite close to hybrid typechecking~\cite{hybtc}. We defer a more detailed discussion of this technique to Section \ref{concl}.

We introduce the static approximation $\mathbf{Stuck}$ for processes that do not return values, but may have side effects. 
\begin{defn}
\mycategory{T}{static approximation}\\
\entry{\cdots}{code}\\
\entry{\mathbf{Stuck}}{stuck process}
\end{defn}
%
We now present rules for stuck-typing. As before, in our first reading of these rules we ignore the side conditions in \textshade[.91]{sharpcorners}{\gdef\outlineboxwidth{0.01}shaded boxes} (which involve the predicate $\ast$). 
\trule{write stuck} identifies code that tries to write to an object whose static trust annotation $\labb$ is higher than the current process label $\labp$. By Invariant (3), the label $\labo$ that protects the object must be at least as high as $\labb$; thus $\labp \sqsubset \labo$ and the code must block at runtime due to access control. For example, let $\mathtt{cmd.exe}$ be of type $\mathbf{Obj}(\_^\top)$ in Example \ref{eg:1}. By \trule{write stuck}, the code $q_1$ is well-typed since $\mathsf{Low} \sqsubset \top$. \trule{un/protect stuck} is similar to \trule{write stuck}; it further identifies code that tries to raise the label of an object beyond the current process label. \trule{escalate stuck} identifies code that tries to raise the current process label. All such processes block at runtime due to access control. 

By \trule{subsumption stuck-I}, processes that are typed under stuck hypotheses are considered stuck as well. For example, this rule combines with \trule{evaluate} to trivially type a continuation $b$ if the intermediate process $a$ is identified as stuck. Finally, by \trule{subsumption stuck-II}, stuck processes can have any type and effect, since they cannot return values.

\subsection{Typing rules for untrusted code}\label{arbit}
Typing must guarantee protection in arbitrary environments. Since the protection guarantee is derived via a type preservation theorem, arbitrary untrusted code needs to be accommodated by the type system. 
We assume that untrusted code runs with a special label $\bot$, introduced into the total order by assuming $\bot \sqsubseteq \lab$ for all $\lab$. 
We now present rules that allow arbitrary interpretation of types at $\bot$. 
\begin{figure}
\cenvvv{Typing rules for untrusted code}{}{
({\bf Typ subsumption $\bot$-I})\vspace{-1mm}
$$
\infer
    {\Gamma, \omega : \mathbf{Obj}(\_^\bot)^{\labt} \vdash_{\labp} a : T}
    {\Gamma, \omega : \mathbf{Obj}(\tau^\bot)^{\labt} \vdash_{\labp} a : T}
$$
~

({\bf Typ subsumption $\bot$-II})\vspace{-1mm}
$$\infer
    {\Gamma, x : \_^\bot \vdash_{\labp} a : T}
    {\Gamma, x : \tau^\bot \vdash_{\labp} a : T}
$$
}
\end{figure}

\noindent
By \trule{subsumption $\bot$-I}, placing the static trust $\bot$ on the contents of an object amounts to assuming any type for those contents as required. By \trule{subsumption $\bot$-II}, a value that has effect $\bot$ may be assumed to have any type as required. These rules provide the necessary flexibility for typing any untrusted code using the other typing rules. On the other hand, arbitrary subtyping with objects can in general be unsound---we now need to be careful when typing trusted code. For example, consider the code
%
$$\omega_2\store{\mathsf{High}}x~\Rsh~\omega_1\store{\mathsf{Low}}\omega_2~\Rsh~[\mathsf{High}]~\eval z {\:!\omega_1}{z := u}$$
A $\mathsf{High}$-process reads the name of an object ($\omega_2$) from a $\mathsf{Low}$-object ($\omega_1$), and then writes $u$ to that object ($\omega_2$). DFI is violated if $\omega_2$ has type $\mathbf{Obj}(\_^{\mathsf{High}})$ and $u$ flows from $\mathsf{Low}$. Unfortunately, it turns out that this code can be typed under process label $\top$ and typing hypotheses
$$\omega_2 : \mathbf{Obj}(\tau_2^{\mathsf{High}})^{\top}\!,~\omega_1 : \mathbf{Obj}(\mathbf{Obj}(\tau_2^{\mathsf{High}})^\bot)^{\top}\!,~x : \tau_2^{\mathsf{High}},~u : \tau_1^{\mathsf{Low}}$$ 
Specifically, the intermediate judgment
%
$$z : \mathbf{Obj}(\tau_2^{\mathsf{High}})^\bot, \dots, u : \tau_1^{\mathsf{Low}} ~\vdash_{\mathsf{High}}~ z := u : \_$$
 can be derived by adjusting the type of $z$ in the typing environment to $\mathbf{Obj}(\tau_1^{\mathsf{Low}})$ with \trule{subsumption $\bot$-II}. 

This source of unsoundness is eliminated if some of the effects in our typing rules are required to be trusted, that is, to be higher than $\bot$. Accordingly we introduce the predicate $\ast$, such that for any label $\lab$, $\ast \lab$ simply means $\lab \sqsupset \bot$. We now revisit the typing rules earlier in the section and focus on the side conditions in \textshade[.91]{sharpcorners}{\gdef\outlineboxwidth{0.01}shaded boxes} (which involve $\ast$). In some of those conditions, we care about trusted effects only if the process label is itself trusted. With these conditions, \trule{write} prevents typechecking the offending write above, since the effect of $z$ in the typing environment is untrusted.

\subsection{Compromise}
The label $\bot$ introduced above is an artificial construct to tolerate a degree of ``anarchy" in the type system. We may want to specify that a certain label (such as $\mathsf{Low}$) acts like $\bot$, \emph{i.e.}, is \emph{compromised}. The typing judgment $\Gamma \vdash_{\labp} a : T~\mathsf{despite}~\labc$ allows us to type arbitrary code $a$ running at a compromised label $\labc$ by assuming that $\labc$ is the same as $\bot$, \emph{i.e.}, by extending the total order with $\labc \sqsubseteq \bot$ (so that all labels that are at most as trusted as $\labc$ collapse to $\bot$). We do not consider labels compromised at runtime (as in Gordon and
Jeffrey's type system for conditional secrecy~\cite{gordonjeffrey05});
however we do not anticipate any technical difficulty in including runtime compromise in our type
system. 

\subsection{Typechecking examples}
We now show some examples of typechecking. 

We begin with the program $p_2$ in Example \ref{eg:2}. Recall that DFI is violated in $p_2$. Suppose that we try to derive the typing judgment 
$$\dots \vdash_\top p_2 : \_ ~~\mathsf{despite}~\mathsf{Low}$$
This amounts to deriving $\dots \vdash_\top p_2 : \_ $ by assuming $\mathsf{Low} \sqsubseteq \bot$. 

As a first step, we apply \trule{new}, \trule{read}, \trule{write}, \trule{pack}, and \trule{evaluate}, directed by syntax, until we have the following typing environment.
%
\begin{eqnarray*}
\Gamma & = & \dots,\\
&&\mathtt{url} : \mathbf{Obj}(\_^\mathsf{Low})^\top, \\
&&\mathtt{setup.exe} : \mathbf{Obj}(\_^\mathsf{Low})^\top, \\
&&\mathtt{binIE} : (\nabla_\mathsf{Low}.~\mathbf{Bin}(\mathbf{Unit}))^\top, \\
&&\mathtt{ie.exe} : \mathbf{Obj}((\nabla_\mathsf{Low}.~\mathbf{Bin}(\mathbf{Unit}))^\top)^\top, \\
&& \mathtt{home} : \mathbf{Obj}(\_^\mathsf{Medium})^\top \\
&& \mathtt{empty} : \mathbf{Unit}^\top
\end{eqnarray*}
The only complication that may arise is in this step is in deriving an intermediate judgment
$$\dots,~ z:\_^\mathsf{Low} ~\vdash_\top~!z : \_$$
Here, we can apply \trule{subsumption $\bot$-II} to adjust the typing hypothesis of $z$ to $\mathbf{Obj}(\_)^\bot$, so that \trule{read} may apply.

After this step, we need to derive a judgment of the form:
$$\Gamma \vdash_\top [\mathsf{High}]~(\dots)~\Rsh~[\mathsf{Medium}]~(\dots)~\Rsh~[\mathsf{Low}]~(\dots)$$
%
Now, we apply \trule{fork}. We first check that the code $[\mathsf{Low}]~(\dots)$ is well-typed. (In fact, untrusted code is always well-typed, as we show in Section \ref{Results}.) The judgment 
$$\Gamma \vdash_\mathsf{Low} \mathtt{home} := \mathtt{empty} : \mathbf{Unit} 
$$
typechecks by \trule{write stuck}. Thus, by \trule{pack} and \trule{evaluate}, we add the following hypothesis to the typing environment.
$$\mathtt{binVirus} : (\nabla_\mathsf{Low}.~\mathbf{Bin}(\mathbf{Unit}))^\mathsf{Low}$$
Let $T_\mathtt{binVirus} = (\nabla_\mathsf{Low}.~\mathbf{Bin}(\mathbf{Unit}))^\mathsf{Low}$. Next, by \trule{new} and \trule{evaluate}, we add the following hypothesis to the typing environment.
$$\mathtt{virus.exe}:\mathbf{Obj}(T_\mathtt{binVirus})^\mathsf{Low}$$
Finally, the judgment 
$$\Gamma, \dots, \mathtt{virus.exe}:\mathbf{Obj}(T_\mathtt{binVirus})^\mathsf{Low} \vdash_\mathsf{Low} \mathtt{url} := \mathtt{virus.exe}$$ 
can be derived by \trule{write}, after massaging the typing hypothesis for $\mathtt{virus.exe}$ to the required $\_^\mathsf{Low}$ by \trule{subsumption $\bot$-II}. 

On the other hand, the process $[\mathsf{High}]~(\dots)$ does not typecheck; as seen above, an intermediate judgment
\begin{equation*}
\Gamma \vdash_\mathsf{High} \mathsf{exec}~\mathtt{setup.exe} : \_
\end{equation*}
cannot be derived, since \trule{execute} does not apply.

To understand this situation further, let us consider some variations where \trule{execute} does apply. Suppose that the code $\mathsf{exec}~z$ is forked in a new process whose label is lowered to $\mathsf{Low}$. Then $p_2$ typechecks. In particular, 
the following judgment can be derived by applying \trule{execute}. 
\begin{equation*}
\Gamma  \vdash_\mathsf{High} [\mathsf{Low}]~\mathsf{exec}~\mathtt{setup.exe} : \_
\end{equation*}
Fortunately, the erasure of $\mathtt{home}$ now blocks by access control at runtime, so DFI is not violated. 

Next, suppose that the static annotation for $\mathtt{setup.exe}$ is $\mathsf{High}$ instead of $\mathsf{Low}$, and $\mathtt{setup.exe}$ is initialized by a process with label $\mathsf{High}$ instead of $\mathsf{Low}$. Then $p_2$ typechecks. In particular, the type of $\mathtt{setup.exe}$ in $\Gamma$ becomes $\mathbf{Obj}(\_^\mathsf{High})$. We need to derive an intermediate judgment 
\begin{equation*}
\Gamma, \dots, x : \_ ~\vdash_\mathsf{Low} \mathtt{setup.exe}:= x : \mathbf{Unit}
\end{equation*}
This judgment can be derived by applying \trule{write stuck} instead of \trule{write}. Fortunately, the overwrite of $\mathtt{setup.exe}$ now blocks by access control at runtime, so DFI is not violated. 

Finally, we sketch how typechecking fails for the violations of DFI described in Section \ref{vista}. 
%
\begin{description}
\item[{\rm (}Write and copy{\rm )}]
  
  Let the type of $\omega$ be $\mathbf{Obj}(\_^{\labb})$, where $\labo \sqsupseteq \labb \sqsupset \labp$. 
  Then the write to $\omega(\labo)$ does not typecheck, since the value to be written is read from $\omega'(\labp)$ and thus has some effect $\labt$ such that $\labt \sqsubseteq \labp$, so that $\labt \sqsubset \labb$.
\item[{\rm (}Copy and execute{\rm )}]
  
  Let the type of $\omega'$ be $\mathbf{Obj}(\_^{\labb'})$. 
  If $\labb' \sqsubseteq \labo$ then the execution of $\omega'(\labp)$ by $q(\labp)$ does not typecheck, since $\labb' \sqsubset \labp$. 
  If $\labb' \sqsupset \labo$ then the write to $\omega'(\labp)$ does not typecheck, since the value to be written is read from $\omega(\labo)$ and thus has some effect $\labt$ such that $\labt \sqsubseteq \labo$, so that $\labt \sqsubset \labb'$.
\item[{\rm (}Unprotect, write, and protect{\rm )}]
  
  Let the type of $\omega$ be $\mathbf{Obj}(\_^{\labb})$, where $\labo \sqsupseteq \labb \sqsupset \labp$. 
  Then the unprotection of $\omega(\labo)$ does not typecheck, since $\labp \sqsubset \labb$.  
\item[{\rm (}Copy, protect, and execute{\rm )}]
  
  Let the type of $\omega'$ be $\mathbf{Obj}(\_^{\labb'})$, where $\labb' \sqsubseteq \labo$. 
  Then the execution of $\omega'(\labp)$ does not typecheck, since $\labb' \sqsubset \labp$.
\end{description}

\section{Properties of typing}\label{Results}
In
this section we show several properties of typing, and prove that DFI is preserved by well-typed code under arbitrary untrusted
environments. All proof details appear in Appendix \ref{proofs}.

We begin with the proposition that untrusted code can always be accommodated by
the type system. 
\begin{definition}[Adversary] A $\labc$-adversary is any process of the form $[\labc]~\_$ that does not contain stores, explicit substitu\-tions, and static trust annotations that are higher than $\labc$.
\end{definition}
%
\begin{proposition}[Adversary completeness]\label{advtyp} Let
  $\Gamma$ be any typing environment and $c$ be any $\labc$-adversary such that $\fv(c) \subseteq
  \mathsf{dom}(\Gamma)$.
Then $\Gamma \vdash_\top c : \_~\mathsf{despite}~\labc$.
\end{proposition}
%
Proposition \ref{advtyp} provides a simple way to quantify over arbitrary environments. By \trule{fork} the composition of a well-typed process with any such environment remains well-typed, and thus enjoys all the properties of typing. 

Next, we present a monotonicity property of typing that is key to decidable and efficient typechecking (Appendix \ref{algo}). 
%
\begin{proposition}[Monotonicity]\label{mono} The following inference rule is admissible.
$$
\infer
    {\Gamma \vdash_{\labp'} f : \tau^{\labt} \\ \Box f \\ \labp \sqsubseteq \labp'}
    {\Gamma \vdash_{\labp} f : \tau^{\labt \sqcap \labp}}
$$
%
\end{proposition}
%
This rule formalizes Invariant (5), and allows inference of ``most general" types for packed code (Appendix \ref{algo}). Further, it implies an intuitive proof principle---code that is proved safe to run with higher privileges remains safe to run with lower privileges, and 
conversely, code that is proved safe against a more powerful adversary remains safe against a less powerful adversary. 

The key property of typing is that it is preserved by structural equivalence and reduction. Preservation depends delicately on the design of the typing rules, relying on the systematic maintenance of typing invariants. 
We write $\Gamma \vdash \sigma$, meaning that ``the substitution environment $\sigma$ is consistent with the typing environment $\Gamma$", if for all $x/u@~\labp \in \sigma$ there exists $T$ such that $x : T \in \Gamma$ and $\Gamma \vdash_{\labp} u : T$.

\begin{theorem}[Preservation]\label{subjred}
  Suppose that $\Gamma \vdash \sigma$ and $\Gamma \vdash_{\labp} a : \_$. Then 
\begin{itemize}
\item if $a \equiv b$ then $\Gamma \vdash_{\labp} b
  : \_$;
\item if $a \action{\labp;\sigma} b$ then $\Gamma \vdash_{\labp} b : \_$.
\end{itemize}
\end{theorem}
%
We now present our formal protection guarantee for well-typed code. 
%
We begin by strengthening the definition of DFI in Section \ref{calc}. In particular, we assume that part of the adversary is known and part of it is unknown. This assumption allows the analysis to exploit any sound typing information that may be obtained from the known part of the adversary. (As a special case, the adversary may be entirely unknown, of course. In this case, we recover Definition \ref{locintdef1}; see below.) Let $\Omega$ be the set of objects that require protection from labels $\lab$ or lower. We let the unknown part of the adversary execute with some process label $\labc$ ($\sqsubseteq \lab$). 
We say that $\Omega$ is protected if no such adversary can write any instance that flows from $\lab$ or lower, to any object in $\Omega$.
%
\begin{definition}[Strong DFI]\label{locintdef} A set of objects $\Omega$ is protected by code $a$ from label $\lab$ despite $\labc$ ($\sqsubseteq \lab$)
  if there is no $\omega \in \Omega$, $\labc$-adversary $c$, substitution environment $\sigma$, and instance $x$ such that $a \Rsh c
  \action{\top,\varnothing}\!\!\!\!{}^\star ~\llctx{\top,\varnothing}{\omega\store{\_} x}{\top,\sigma}$ and $x\stackrel\sigma\blacktriangledown \lab$.
\end{definition}
%
For example, we may want to prove that some code protects a set of $\mathsf{High}$-objects from $\mathsf{Medium}$ despite (the compromised label) $\mathsf{Low}$; then we need to show that no instance may flow from $\mathsf{Medium}$ or lower to any of those $\mathsf{High}$-objects under any $\mathsf{Low}$-adversary.

We pick objects that require protection based on their types and effects in the typing environment.
\begin{definition}[Trusted objects] The set of objects whose contents are trusted beyond the label $\lab$ in the typing environment $\Gamma$ is $\{\omega~|~\omega :
  \mathbf{Obj}(\_^{\labb})^{\labt} \in \Gamma\mbox{ and }\labb \sqcap \labt \sqsupset
  \lab\}$.
\end{definition}
%
Suppose that in some typing environment, $\Omega$ is the set of objects whose contents are trusted beyond label $\lab$, and $\labc$ ($\sqsubseteq \lab$) is compromised; we guarantee that $\Omega$ is protected by any well-typed code from $\lab$ despite $\labc$.
\begin{theorem}[Enforcement of strong DFI]\label{mainthm} Let
  $\Omega$ be the set of objects whose contents are trusted beyond $\lab$ in $\Gamma$. Suppose that $\Gamma
  \vdash_\top a : \_~\mathsf{despite}~\labc$, where $\labc \sqsubseteq \lab$. Then $a$
  protects $\Omega$ from $\lab$ despite $\labc$. 
\end{theorem}
%
In the special case where the adversary is entirely unknown, we simply consider $\lab$ and $\labc$ to be the same label. 

The type system further enforces DFI for new objects, as can be verified by applying Theorem \ref{subjred}, \trule{substitute}, and Theorem \ref{mainthm}. 
%
Finally, the type system suggests a sound runtime optimization:
whenever a well-typed process executes packed code in a trusted context, the current process label is already appropriately lowered for execution. 
%
\begin{theorem}[Redundancy of execution control]\label{optim} Suppose that $\Gamma \vdash_{\top} a : \_ ~\mathsf{despite}~\labc$ and  
$a \action{\top;\varnothing}\!\!\!\!{}^\star ~\llctx{\top;\varnothing}{\fork{\omega \store{\labo} \_} \mathsf{exec}\:\omega'}{\labp;\sigma}$ such that $\omega\stackrel\sigma= \omega'$ and $\labp \sqsupset\labc$. Then $\labp \sqsubseteq \labo$. 
\end{theorem}
It follows that the rule (\rrule{execute}) can be safely optimized as follows.
$$\infer
    {\omega\stackrel\sigma= \omega' \\ \mathsf{pack}(f) \in \sigma(x)}
    {\fork{\omega \store{\labo} x}\mathsf{exec}~\omega' \action{\labp;\sigma} \fork{\omega \store{\labo} x}f}
$$
This optimization should not be surprising. Lowering the process label for execution aims to prevent trusted code from executing untrusted code in trusted contexts; our core static discipline on trusted code effectively subsumes this runtime control. 
 On the other hand, write-access control cannot be eliminated by any discipline on trusted code, since that control is required to restrict untrusted code. 

Lastly, typechecking can be efficiently mechanized thanks to Proposition \ref{mono} and our syntactic restriction on nested packing. 
A typechecking algorithm is outlined in Appendix \ref{algo}.
\begin{theorem}[Typechecking]\label{decide} Given a typing environment $\Gamma$ and code $a$ with $\mathbb L$ distinct labels, the problem of whether there exists $T$ such that $\Gamma \vdash_\top a : T$, is decidable in time $\mathcal O(\mathbb L |a|)$, where $|a|$ is the size of $a$.
\end{theorem}
\section{Limitations, related work, and discussion}\label{concl}

In this paper we formalize DFI---a multi-level integrity property based on explicit flows---and present a type system that can efficiently enforce DFI in a language that simulates Windows Vista's security environment. 

Not surprisingly, our type system is only a conservative technique to enforce
DFI---while every program that typechecks is guaranteed to
satisfy DFI (as stated in Theorem~\ref{mainthm}), 
well-typedness is not necessary for DFI. 

By design, our analysis is control-insensitive---it does not track implicit flows. In many applications, implicit flows are of serious concern. It remains possible to extend our analysis to account for such flows, following the ideas of \cite{vis,robustdeclass,myers04enforcing,lipopl}. However, we believe that it is more practical to enforce a weaker property like DFI at the level of an operating system, and enforce stronger, control-sensitive properties like noninterference at the level of the application, with specific assumptions. 

Our core security calculus is simplified,
although we take care to include all aspects that require conceptual modeling 
for reasoning
about DFI.
In particular, we model threads, mutable references, binaries, and data and code pointers; other features of x86 binaries, 
such as recursion, control flow, and parameterized procedures, can be encoded in the core calculus.
We also model all details of Windows Vista that are relevant for mandatory integrity control with dynamic labels. On the other hand, we do not model details 
such as discretionary access control, file virtualization, and secure authorization of privilege escalation \cite{bppvista}, which can improve the precision of our analysis.
Building a typechecker that works at the level of x86 binaries and handles
all details of Windows Vista requires more work. At the same time, we believe that our analysis can be applied to more concrete programming models by translation. 

Our work is closely related to that of Tse and
Zdancewic~\cite{runtimeprin} and Zheng and Myers \cite{dynseclab} on
noninterference in lambda calculi with dynamic security levels. While Tse and Zdancewic do not consider mutable references in their language, it is possible to encode the sequential fragment of our calculus in the language of Zheng and Myers; however, well-typed programs in that fragment that rely on access control for DFI do not remain well-typed via such an encoding. Specifically, any restrictive access check for integrity in the presence of dynamically changing labels seems to let the adversary influence trusted computations in their system, violating noninterference~\cite{zheng}. 

%
%


 Noninterference is known to be problematic for concurrent languages. In this context, Zdancewic and Myers study the notion of observational determinism \cite{obsdetconc}; Abadi, Hennessy and Riely, and others study information flow using testing equivalence~\cite{abadi99secrecy,inflowresacc}; and Boudol and Castellani, Honda and Yoshida, and others use stronger notions based on observational equivalence~\cite{boudolcastellani,linearpi}. So\-phisticated techniques that involve linearity, race analysis, behavior types, and liveness analysis also appear in the literature~\cite{linearpi,obsdetconc,inflowresacc,kobayashi05}. While most of these techniques are developed in the setting of the pi calculus, other works consider distributed and higher-order settings to study mobile code~\cite{safedpi,depHOMP,envbisim} (as in this work).

DFI being a safety property \cite{alpernschneider} gets around some of the difficulties posed by noninterference. A related approach guides the design of the operating systems  Asbestos \cite{asbestos} and HiStar \cite{histar}, and dates back to the Clark-Wilson approach to security in commercial computer systems \cite{clark-wilson,shankar}. In comparison with generic models of trace-based integrity that appear in protocol analysis, such as correspondence assertions \cite{typecorr,fournetGM05}, our integrity model is far more specialized; as a consequence, our type system requires far less annotations than type systems for proving correspondence assertions. 

Our definition of
DFI relies on an operational semantics based on
explicit substitution. Explicit substitution, as introduced by Abadi \emph{et al.} \cite{abadi90explicit}, has been primarily applied to study the
correctness of abstract machines for programming languages (whose semantics rely on substitution as a rather inefficient
meta-operation), and in proof environments. It also appears in the applied pi
calculus \cite{appliedpi} to facilitate an elegant formulation of
indistinguishability for security analysis. However, we seem to be
the first to use explicit substitutions to track explicit flows in
a concurrent language. 
Previously, dependency analysis
\cite{levylabels,cachedep} has been applied to information-flow analysis
\cite{dcc,pottier00information,ZM02}. These analyses track stronger dependencies than those induced by explicit flows; in particular, the dependencies are sensitive to control flows. In contrast, the use of explicit substitutions to track explicit flows seems rather obvious and appropriate in hindsight. We believe that this technique should be useful in other contexts as well. 

Our analysis manifests a genuine interplay between static typing and dynamic access control for runtime protection. We seem to be the first to study this interaction in a concurrent system with dynamic labels for multi-level integrity. This approach of combining static and dynamic protection mechanisms is reflected in previous work, \emph{e.g.}, on typing for noninterference in a Java-like language with stack inspection and other extensions \cite{banerjee03using,pistoia}, for noninterference in lambda calculi with runtime principals and dynamic labels~\cite{runtimeprin,dynseclab}, and for secrecy in concurrent storage calculi with discretionary access control mechanisms~\cite{sectypfac,ChaudhuriConcur06}. A verification technique based on this approach is developed by Flanagan~\cite{hybtc} for a lambda calculus with arbitrary base refinement types. In these studies and ours, dynamic checks complement static analysis where possible or as required, so that safety violations that are not caught statically are always caught at runtime. Moreover, static typing sometimes subsumes certain dynamic checks (as in our analysis), suggesting sound runtime optimizations. This approach is reflected in previous work on static access control \cite{inflowresacc,pottier-skalka-smith-toplas05,sumii}. 

In most real-world systems, striking the right balance between security and practice is a delicate task that is never far from controversy. It is reassuring to discover that perhaps, such a balance can be enforced formally in a contemporary operating system, and\linebreak possibly improved in future ones.

\paragraph{\em Acknowledgments}
We wish to thank Mart\'in Abadi, Steve Zdan\-cewic, Pavol \v{C}ern\'y,  and several anonymous reviewers for their comments on an earlier draft of this paper. We also wish to thank 
Karthik Bhargavan, Cormac Flanagan, and Lantian Zheng for various discussions on this work. 

Avik Chaudhuri's work was supported by Microsoft Research India and the National Science
Foundation under Grants CCR-0208800 and CCF-0524078.

\bibliographystyle{abbrv}
\bibliography{vista}

\begin{thebibliography}{10}

\bibitem{abadi99secrecy}
M.~Abadi.
\newblock Secrecy by typing in security protocols.
\newblock {\em Journal of the ACM}, 46(5):749--786, 1999.

\bibitem{dcc}
M.~Abadi, A.~Banerjee, N.~Heintze, and J.~G. Riecke.
\newblock A core calculus of dependency.
\newblock In {\em POPL'99: Principles of Programming Languages}, pages
  147--160. ACM, 1999.

\bibitem{sectyplog}
M.~Abadi and B.~Blanchet.
\newblock Analyzing security protocols with secrecy types and logic programs.
\newblock In {\em POPL'02: Principles of Programming Languages}, pages 33--44.
  ACM, 2002.

\bibitem{abadi90explicit}
M.~Abadi, L.~Cardelli, P.-L. Curien, and J.-J. L\'{e}vy.
\newblock Explicit substitutions.
\newblock In {\em POPL'90: Principles of Programming Languages}, pages 31--46.
  ACM, 1990.

\bibitem{appliedpi}
M.~Abadi and C.~Fournet.
\newblock Mobile values, new names, and secure communication.
\newblock In {\em POPL'01: Principles of Programming Languages}, pages
  104--115. ACM, 2001.

\bibitem{cachedep}
M.~Abadi, B.~Lampson, and J.-J. L\'{e}vy.
\newblock Analysis and caching of dependencies.
\newblock In {\em ICFP'96: Functional Programming}, pages 83--91. ACM, 1996.

\bibitem{alpernschneider}
B.~Alpern and F.~B. Schneider.
\newblock Defining liveness.
\newblock {\em Information Processing Letters}, 21(5):181--185, 1985.

\bibitem{banerjee03using}
A.~Banerjee and D.~Naumann.
\newblock Using access control for secure information flow in a {J}ava-like
  language.
\newblock In {\em CSFW'03: Computer Security Foundations Workshop}, pages
  155--169. IEEE, 2003.

\bibitem{biba}
K.~J. Biba.
\newblock Integrity considerations for secure computer systems.
\newblock Technical Report TR-3153, MITRE Corporation, 1977.

\bibitem{boudolcastellani}
G.~Boudol and I.~Castellani.
\newblock Noninterference for concurrent programs and thread systems.
\newblock {\em Theoretical Computer Science}, 281(1-2):109--130, 2002.

\bibitem{secgp}
L.~Cardelli, G.~Ghelli, and A.~D. Gordon.
\newblock Secrecy and group creation.
\newblock {\em Information and Computation}, 196(2):127--155, 2005.

\bibitem{castro}
M.~Castro, M.~Costa, and T.~Harris.
\newblock Securing software by enforcing data-flow integrity.
\newblock In {\em OSDI'06: Operating Systems Design and Implementation}, pages
  147--160. USENIX, 2006.

\bibitem{ChaudhuriConcur06}
A.~Chaudhuri.
\newblock Dynamic access control in a concurrent object calculus.
\newblock In {\em CONCUR'06: Concurrency Theory}, pages 263--278. Springer,
  2006.

\bibitem{sectypfac}
A.~Chaudhuri and M.~Abadi.
\newblock Secrecy by typing and file-access control.
\newblock In {\em CSFW'06: Computer Security Foundations Workshop}, pages
  112--123. IEEE, 2006.

\bibitem{clark-wilson}
D.~D. Clark and D.~R. Wilson.
\newblock A comparison of commercial and military computer security policies.
\newblock In {\em SP'87: Symposium on Security and Privacy}, pages 184--194.
  IEEE, 1987.

\bibitem{dytan}
J.~Clause, W.~Li, and A.~Orso.
\newblock Dytan: a generic dynamic taint analysis framework.
\newblock In {\em ISSTA'07: International Symposium on Software Testing and
  Analysis}, pages 196--206. ACM, 2007.

\bibitem{symantec}
M.~Conover.
\newblock Analysis of the windows vista security model.
\newblock Available at \url{www.symantec.com/avcenter/reference/
  Windows_Vista_Security_Model_Analysis.pdf}.

\bibitem{denningcert}
D.~E. Denning and P.~J. Denning.
\newblock Certification of programs for secure information flow.
\newblock {\em Communications of the ACM}, 20(7):504--513, 1977.

\bibitem{asbestos}
P.~Efstathopoulos, M.~Krohn, S.~VanDeBogart, C.~Frey, D.~Ziegler, E.~Kohler,
  D.~Mazi\`{e}res, F.~Kaashoek, and R.~Morris.
\newblock Labels and event processes in the {A}sbestos operating system.
\newblock In {\em SOSP'05: Symposium on Operating Systems Principles}, pages
  17--30. ACM, 2005.

\bibitem{evalcontext}
M.~Felleisen.
\newblock The theory and practice of first-class prompts.
\newblock In {\em POPL'88: Principles of Programming Languages}, pages
  180--190. ACM, 1988.

\bibitem{hybtc}
C.~Flanagan.
\newblock Hybrid type checking.
\newblock In {\em POPL'06: Principles of Programming Languages}, pages
  245--256. ACM, 2006.

\bibitem{fournetGM05}
C.~Fournet, A.~D. Gordon, and S.~Maffeis.
\newblock A type discipline for authorization policies.
\newblock In {\em ESOP'05: European Symposium on Programming}, pages 141--156.
  Springer, 2005.

\bibitem{nonintf}
J.~A. Goguen and J.~Meseguer.
\newblock Security policies and security models.
\newblock In {\em SP'82: Symposium on Security and Privacy}, pages 11--20.
  IEEE, 1982.

\bibitem{gordon98concurrent}
A.~D. Gordon and P.~D. Hankin.
\newblock A concurrent object calculus: Reduction and typing.
\newblock In {\em {HLCL}'98: High-Level Concurrent Languages}, pages 248--264.
  Elsevier, 1998.

\bibitem{typecorr}
A.~D. Gordon and A.~Jeffrey.
\newblock Typing correspondence assertions for communication protocols.
\newblock {\em Theoretical Computer Science}, 300(1-3):379--409, 2003.

\bibitem{gordonjeffrey05}
A.~D. Gordon and A.~Jeffrey.
\newblock Secrecy despite compromise: Types, cryptography, and the pi-calculus.
\newblock In {\em CONCUR'05: Concurrency Theory}, pages 186--201. Springer,
  2005.

\bibitem{safedpi}
M.~Hennessy, J.~Rathke, and N.~Yoshida.
\newblock Safe{D}pi: A language for controlling mobile code.
\newblock {\em Acta Informatica}, 42(4-5):227--290, 2005.

\bibitem{inflowresacc}
M.~Hennessy and J.~Riely.
\newblock Information flow vs. resource access in the asynchronous pi-calculus.
\newblock {\em ACM Transactions on Programming Languages and Systems},
  24(5):566--591, 2002.

\bibitem{linearpi}
K.~Honda and N.~Yoshida.
\newblock A uniform type structure for secure information flow.
\newblock In {\em POPL'02: Principles of Programming Languages}, pages 81--92.
  ACM, 2002.

\bibitem{sumii}
D.~Hoshina, E.~Sumii, and A.~Yonezawa.
\newblock A typed process calculus for fine-grained resource access control in
  distributed computation.
\newblock In {\em TACS'01: Theoretical Aspects of Computer Software}, pages
  64--81. Springer, 2001.

\bibitem{bppvista}
M.~Howard and D.~LeBlanc.
\newblock {\em Writing Secure Code for Windows Vista}.
\newblock Microsoft Press, 2007.

\bibitem{kobayashi05}
N.~Kobayashi.
\newblock Type-based information flow analysis for the pi-calculus.
\newblock {\em Acta Informatica}, 42(4-5):291--347, 2005.

\bibitem{lamport77}
L.~Lamport.
\newblock Proving the correctness of multiprocess programs.
\newblock {\em IEEE Transactions on Software Engineering}, 3(2):125--143, 1977.

\bibitem{protection}
B.~W. Lampson.
\newblock Protection.
\newblock {\em ACM Operating Systems Review}, 8(1):18--24, Jan 1974.

\bibitem{levylabels}
J.-J. L\'{e}vy.
\newblock {\em R\'{e}ductions correctes et optimales dans le lambda-calcul}.
\newblock PhD thesis, Universit\'{e} Paris 7, 1978.

\bibitem{lipopl}
P.~Li and S.~Zdancewic.
\newblock Downgrading policies and relaxed noninterference.
\newblock In {\em POPL'05: Principles of Programming Languages}, pages
  158--170. ACM, 2005.

\bibitem{perl}
L.ÊWall, T.ÊChristiansen, and R.ÊSchwartz.
\newblock {\em Programming Perl}.
\newblock O'Reilly, 1996.

\bibitem{myers04enforcing}
A.~Myers, A.~Sabelfeld, and S.~Zdancewic.
\newblock Enforcing robust declassification.
\newblock In {\em CSFW'04: Computer Security Foundations Workshop}, pages
  172--186. IEEE, 2004.

\bibitem{pcc}
G.~C. Necula.
\newblock Proof-carrying code.
\newblock In {\em POPL'97: Principles of Programming Languages}, pages
  106--119. ACM, 1997.

\bibitem{pistoia}
M.~Pistoia, A.~Banerjee, and D.~A. Naumann.
\newblock Beyond stack inspection: A unified access-control and
  information-flow security model.
\newblock In {\em SP'07: Symposium on Security and Privacy}, pages 149--163.
  IEEE, 2007.

\bibitem{pottier00information}
F.~Pottier and S.~Conchon.
\newblock Information flow inference for free.
\newblock In {\em ICFP'00: Functional Programming}, pages 46--57. ACM, 2000.

\bibitem{pottier-skalka-smith-toplas05}
F.~Pottier, C.~Skalka, and S.~Smith.
\newblock A systematic approach to static access control.
\newblock {\em ACM Transactions on Programming Languages and Systems},
  27(2):344--382, 2005.

\bibitem{uac}
M.~Russinovich.
\newblock {\em Inside Windows Vista User Access Control}.
\newblock Microsoft Technet Magazine, June 2007.
\newblock Available at
  \url{http://www.microsoft.com/technet/technetmag/issues/2007/06/UAC/}.

\bibitem{sabelfeld}
A.~Sabelfeld and A.~Myers.
\newblock Language-based information-flow security.
\newblock {\em IEEE Journal on Selected Areas in Communications}, 21(1), 2003.

\bibitem{envbisim}
D.~Sangiorgi, N.~Kobayashi, and E.~Sumii.
\newblock Environmental bisimulations for higher-order languages.
\newblock In {\em LICS'07: Logic in Computer Science}, pages 293--302. IEEE,
  2007.

\bibitem{shankar}
U.~Shankar, T.~Jaeger, and R.~Sailer.
\newblock Toward automated information-flow integrity verification for
  security-critical applications.
\newblock In {\em NDSS'06: Network and Distributed System Security Symposium}.
  ISOC, 2006.

\bibitem{suh}
G.~E. Suh, J.~W. Lee, D.~Zhang, and S.~Devadas.
\newblock Secure program execution via dynamic information flow tracking.
\newblock In {\em ASPLOS'04: Architectural Support for Programming Languages
  and Operating Systems}, pages 85--96. ACM, 2004.

\bibitem{runtimeprin}
S.~Tse and S.~Zdancewic.
\newblock Run-time principals in information-flow type systems.
\newblock In {\em SP'04: Symposium on Security and Privacy}, pages 179--193.
  IEEE, 2004.

\bibitem{vogt}
P.~Vogt, F.~Nentwich, N.~Jovanovic, C.~Kruegel, E.~Kirda, and G.~Vigna.
\newblock Cross site scripting prevention with dynamic data tainting and static
  analysis.
\newblock In {\em NDSS'07: Network and Distributed System Security Symposium}.
  ISOC, 2007.

\bibitem{vis}
D.~Volpano, C.~Irvine, and G.~Smith.
\newblock A sound type system for secure flow analysis.
\newblock {\em Journal of Computer Security}, 4(2-3):167--187, 1996.

\bibitem{blame}
P.~Wadler and R.~B. Findler.
\newblock Well-typed programs can't be blamed.
\newblock In {\em Scheme'07: Workshop on Scheme and Functional Programming},
  2007.

\bibitem{panorama}
H.~Yin, D.~Song, M.~Egele, C.~Kruegel, and E.~Kirda.
\newblock Panorama: capturing system-wide information flow for malware
  detection and analysis.
\newblock In {\em CCS'07: Computer and Communications Security}, pages
  116--127. ACM, 2007.

\bibitem{depHOMP}
N.~Yoshida.
\newblock Channel dependent types for higher-order mobile processes.
\newblock In {\em POPL'04: Principles of Programming Languages}, pages
  147--160. ACM, 2004.

\bibitem{robustdeclass}
S.~Zdancewic and A.~C. Myers.
\newblock Robust declassification.
\newblock In {\em CSFW'01: Computer Security Foundations Workshop}, pages
  5--16. IEEE, 2001.

\bibitem{ZM02}
S.~Zdancewic and A.~C. Myers.
\newblock Secure information flow via linear continuations.
\newblock {\em Higher Order and Symbolic Computation}, 15(2/3):209--234, 2002.

\bibitem{obsdetconc}
S.~Zdancewic and A.~C. Myers.
\newblock Observational determinism for concurrent program security.
\newblock In {\em CSFW'03: Computer Security Foundations Workshop}, pages
  29--43. IEEE, 2003.

\bibitem{histar}
N.~Zeldovich, S.~Boyd-Wickizer, E.~Kohler, and D.~Mazi\`{e}res.
\newblock Making information flow explicit in {H}i{S}tar.
\newblock In {\em OSDI'06: Operating Systems Design and Implementation}, pages
  19--19. USENIX, 2006.

\bibitem{zheng}
L.~Zheng.
\newblock Personal communication, July 2007.

\bibitem{dynseclab}
L.~Zheng and A.~Myers.
\newblock Dynamic security labels and noninterference.
\newblock In {\em FAST'04: Formal Aspects in Security and Trust}, pages 27--40.
  Springer, 2004.

\end{thebibliography}


\appendix








\section*{Appendix}
In this appendix, we provide some additional material that may benefit the reader. First, we detail proofs of our results on typing (Appendix \ref{proofs}). Next, we outline an efficient typechecking algorithm (Appendix \ref{algo}).

\section{Proofs}\label{proofs}
\noindent
In this section we outline proofs of the results in Section \ref{Results}. 

$ $\\
{\bf Restatement of Proposition \ref{advtyp}} (Adversary completeness) {\em Let
  $\Gamma$ be any typing environment and $e$ be any $\labc$-adversary  such that $\fv(e) \subseteq
  \mathsf{dom}(\Gamma)$.
Then $\Gamma \vdash_\top e : \_~\mathsf{despite}~\labc$.
}
%
\begin{proof} We prove typability by induction on the structure of processes.
\begin{itemize}
\item $e \equiv x$ where $u$ is a variable.\\
\\
Then $x \in \dom(\Gamma)$.\\
By \trule{value} $\Gamma \vdash_{\labc} x :_\_$.\\
\item $e \equiv \mathsf{new}(x\mbox{ \# }\labb)$.\\
\\
By I.H. $\Gamma \vdash_{\labc} x : \tau^{\labt}$\\
Then $\labb \sqsubseteq \labc \sqsubseteq \bot \sqsubseteq \labt$.\\
By \trule{new} $\Gamma \vdash_{\labc} \mathsf{new}(x\mbox{ \# }\labb) : \_$.\\
\item $e \equiv \langle\labo\rangle~\omega$.\\
\\
By I.H. $\Gamma \vdash_{\labc} \omega : \_$.\\
So by \trule{value} $\omega : \tau^{\labt} \in \Gamma$.
\begin{description}
\item[Case] $\ast \labt$ and $\tau$ is not of the form $\mathbf{Obj}(\_)$.\\
By \trule{bogus stuck-I} $\Gamma \vdash_{\labc} \langle\labo\rangle~\omega : \_$.
\item[Case] $\ast \labt$, $\tau = \mathbf{Obj}(\_^{\labb})$, and $\labc \sqsubset \labb \sqcup \labo$.\\
By \trule{un/protect stuck} $\Gamma \vdash_{\labc} \langle\labo\rangle~\omega : \_$.
\item[Case] $\ast \labt$, $\tau = \mathbf{Obj}(\_^{\labb})$, and $\bot \sqsubseteq \labb \sqcup \labo \sqsubseteq \labc = \bot$.\\
Then $\labb \sqsubseteq \labo$.\\
By \trule{value} and \trule{un/protect} \\
$~~~$ $\Gamma \vdash_{\labc} \langle\labo\rangle~\omega : \_$.
\item[Case] $\labt = \bot$.\\
By \trule{subsumption $\bot$-II} \\
$~~~$ $\tau = \mathbf{Obj}(\_^{\labb})$ such that $\labb \sqsubseteq \labo$.\\
By \trule{value} and \trule{un/protect} \\
$~~~$ $\Gamma \vdash_{\labc} \langle\labo\rangle~\omega : \_$.
\end{description}
$ $
%
\item $e \equiv ~!\omega$.\\
\\
By I.H. $\Gamma \vdash_{\labc} \omega : \_$.\\
So by \trule{value} $\omega : \tau^{\labt} \in \Gamma$.
\begin{description}
\item[Case] $\ast \labt$ and $\tau$ is not of the form $\mathbf{Obj}(\_)$.\\
By \trule{bogus stuck-I} $\Gamma \vdash_{\labc} ~!\omega : \_$.
\item[Case] $\ast \labt$ and $\tau = \mathbf{Obj}(\_)$.\\
By \trule{read} $\Gamma \vdash_{\labc} ~!\omega : \_$.
\item[Case] $\labt = \bot$.\\
By \trule{subsumption $\bot$-II} $\tau = \mathbf{Obj}(\_)$.\\
By \trule{read} $\Gamma \vdash_{\labc} ~!\omega : \_$.
\end{description}
$ $
%
\item $e \equiv \omega := x$.\\
\\
By I.H. $\Gamma \vdash_{\labc} \omega : \_$ and $\Gamma \vdash_{\labc} x : \tau_1^{\labt'}$.\\
So by \trule{value} $\omega : \tau^{\labt} \in \Gamma$.
\begin{description}
\item[Case] $\ast \labt$ and $\tau$ is not of the form $\mathbf{Obj}(\_)$.\\
By \trule{bogus stuck-I} $\Gamma \vdash_{\labc} \omega := x : \_$.
\item[Case] $\ast \labt$, $\tau = \mathbf{Obj}(\_^{\labb})$, and $\labc \sqsubset \labb$.\\
By \trule{write stuck} $\Gamma \vdash_{\labc} \omega := x : \_$.
\item[Case] $\ast \labt$, $\tau = \mathbf{Obj}(\tau_1^{\labb})$, and $\bot \sqsubseteq \labb \sqsubseteq \labc = \bot$.\\
Then $\labb \sqsubseteq \labt'$.\\
By \trule{value} and \trule{write} $\Gamma \vdash_{\labc} \omega := x : \_$.
\item[Case] $\labt = \bot$.\\
By \trule{subsumption $\bot$-II} \\
$~~~$ $\tau = \mathbf{Obj}(\tau_1^{\labb})$ such that $\labb \sqsubseteq \labt'$.\\
By \trule{value} and \trule{write} $\Gamma \vdash_{\labc} \omega := x : \_$.
\end{description}
$ $
%
\item $e \equiv \mathsf{pack}(f)$.\\
\\
By I.H. $\Gamma \vdash_{\labc} f : T$.\\
By \trule{pack} $\Gamma \vdash_{\labc} \mathsf{pack}(f) : \_$.\\
\item $e \equiv \mathsf{exec}~\omega$.\\
\\
By I.H. $\Gamma \vdash_{\labc} \omega : \_$, so by \trule{value} $\omega : \tau^{\labt} \in \Gamma$.
\begin{description}
\item[Case] $\ast \labt$ and $\tau$ is not of the form $\mathbf{Obj}(\_)$.\\
By \trule{bogus stuck-I} $\Gamma \vdash_{\labc} \mathsf{exec}~\omega : \_$.
\item[Case] $\tau = \mathbf{Obj}(\tau_1^{\labb})$, $\ast \labt$, \\
$~~~$ and $\tau_1$ is not of the form $\nabla_\_.~\mathbf{Bin}(\_)$.\\
By \trule{bogus stuck-II} $\Gamma \vdash_{\labc} \mathsf{exec}~\omega : \_$.
\item[Case] $\tau = \mathbf{Obj}(\tau_1^{\labb})$, $\ast \labt$, and $\tau_1 = \nabla_{\labp}.~\mathbf{Bin}(\_)$.\\
Then $\labc = \bot \sqsubseteq \labp \sqcap \labb$.\\
By \trule{execute} $\Gamma \vdash_{\labc} \mathsf{exec}~\omega : \_$.
\item[Case] $\labt = \bot$.\\
By \trule{subsumption $\bot$-II} \\
$~~~$ $\tau = \mathbf{Obj}(\tau_1^{\labb})$ and $\tau_1 = \nabla_{\labp}.~\mathbf{Bin}(\_)$\\
$~~~$ such that $\labc = \bot \sqsubseteq \labp \sqcap \labb$.\\
By \trule{execute} $\Gamma \vdash_{\labc} \mathsf{exec}~\omega : \_$.
\item[Case] $\ast \labt$, $\tau = \mathbf{Obj}(\tau_1^{\labb})$, and $\labb = \bot$.\\
By \trule{subsumption $\bot$-I}\\
$~~~$ $\tau_1 = \nabla_{\labp}.~\mathbf{Bin}(\_)$ such that $\labc = \bot \sqsubseteq \labp \sqcap \labb$.\\
By \trule{execute} $\Gamma \vdash_{\labc} \mathsf{exec}~\omega : \_$.
\end{description}
$ $
%
\item $e \equiv [\labp]~a$.\\
\\
If $\labp \sqsupset \labc$ then by \trule{escalate} $\Gamma \vdash_{\labc} [\labp]~a : \_$.\\
Otherwise by I.H. $\Gamma \vdash_{\labp} a : \_$.\\
$~~~$ By \trule{limit} $\Gamma \vdash_{\labc} [\labp]~a : \_$.\\
\item $e \equiv \eval x a b$.\\
\\
By I.H. $\Gamma \vdash_{\labc} a : T$\\
$~~~$ and  $\Gamma, x: T \vdash_{\labc} b : T'$.\\
By \trule{evaluate} $\Gamma \vdash_{\labc} \eval x a b : \_$.\\
\item $e \equiv \fork a b$.\\
\\
By I.H. $\Gamma \vdash_{\labc} a : \_$\\
$~~~$ and  $\Gamma \vdash_{\labc} b : T$.\\
By \trule{fork} $\Gamma \vdash_{\labc} \fork a b : \_$.  \qedhere
%
\end{itemize}
\end{proof}
%
%
%
%
$ $\\
{\bf Restatement of Proposition \ref{mono}} (Monotonicity) {\em
The following typing rule is admissible.
$$
\infer
    {\Gamma \vdash_{\labp'} f : \tau^{\labt} \\ \Box f \\ \labp \sqsubset \labp'}
    {\Gamma \vdash_{\labp} f : \tau^{\labt \sqcap \labp}}
$$
%
}
\begin{proof} We proceed by induction on the structure of derivations.\\
\\
Suppose that $\labp' \sqsubset \labp$.
\begin{description}
\item[Case (Typ variable)] 
$$\infer
    {x : \tau^{\labt} \in \Gamma}
    {\Gamma \vdash_{\labp} x : \tau^{\labt \sqcap \labp}}
$$
By \trule{value} $\Gamma \vdash_{\labp} x : \tau^{\labt \sqcap \labp'}$.\\
Here $\labt \sqcap \labp' = \labt \sqcap \labp \sqcap \labp'$.\\
%
\item[Case (Typ new)]
$$\infer
    {\Gamma \vdash_{\labp} x : \tau^{\labt} \\ \labb \sqsubseteq \labt}
    {\Gamma \vdash_{\labp} \mathsf{new}(x\mbox{ \# }\labb) : \mathbf{Obj}(\tau^{\labb})^{\labp}}
$$
By I.H. $\Gamma \vdash_{\labp'} x : \tau^{\labt \sqcap \labp'}$\\
Then $\labb \sqsubseteq \labt \sqcap \labp'$.\\
By \trule{new} $\Gamma \vdash_{\labp'} \mathsf{new}(x\mbox{ \# }\labb) : \mathbf{Obj}(\tau^{\labb})^{\labp'}$.\\
Here $\labp' = \labp \sqcap \labp'$.\\
\item[Case (Typ fork)]
$$\infer
    {\Gamma \vdash_{\labp} a : \_ \\
    \Gamma \vdash_{\labp} b : T}
    {\Gamma \vdash_{\labp} \fork a b : T}
$$
Let $T = \tau^{\labt}$.\\
By I.H. $\Gamma \vdash_{\labp'} a : \_ $ and $\Gamma \vdash_{\labp'} b : \tau^{\labt \sqcap \labp'}$.\\
By \trule{fork} $\Gamma \vdash_{\labp'} a \Rsh b :  \tau^{\labt \sqcap \labp'}$.\\
\item[Case (Typ store)]
$$\infer
    {\{\omega : \mathbf{Obj}(\tau^{\labb})^\_, x : \tau^{\labt}\} \subseteq \Gamma \\
    \labb \sqsubseteq \labo \sqcap \labt}
    {\Gamma \vdash_{\labp} \omega \store{\labo} x : \_^{\labp}}
$$
By \trule{store} $\Gamma \vdash_{\labp'} \omega \store{\labo} x : \_^{\labp'}$.\\
Here $\labp' = \labp \sqcap \labp'$.\\
\item[Case (Typ un/protect)]
$$\inferrule*[right= ~\rem{\ast\labp\Rightarrow\ast\labt}]
 {\Gamma \vdash_{\labp} \omega :
    \mathbf{Obj}(\tau^{\labb})^{\labt} \\
    \labb \sqsubseteq \labo}
    {\Gamma \vdash_{\labp} \langle\labo\rangle\: \omega : \mathbf{Unit}^{\labp}}
$$
By I.H. $\Gamma \vdash_{\labp'} \omega :
    \mathbf{Obj}(\tau^{\labb})^{\labt \sqcap \labp'}$\\
and if $\ast \labp'$ then $\ast\labp$, then $\ast \labt$, and then $\ast (\labt \sqcap \labp')$.\\
By \trule{un/protect} $\Gamma \vdash_{\labp'} \langle\labo\rangle\: \omega : \mathbf{Unit}^{\labp'}$.\\
%
\item[Case (Typ write)]
$$\inferrule*[right= ~\rem{\ast\labp \Rightarrow \ast\labt} ]
    {\Gamma \vdash_{\labp} \omega :
    \mathbf{Obj}(\tau^{\labb})^{\labt} \\
    \Gamma \vdash_{\labp} x : \tau^{\labt'} \\
     \labb \sqsubseteq \labt' }
    {\Gamma \vdash_{\labp} \omega := x : \mathbf{Unit}^{\labp}}
$$
By I.H. $\Gamma \vdash_{\labp'} \omega :
    \mathbf{Obj}(\tau^{\labb})^{\labt \sqcap \labp'}$ and $\Gamma \vdash_{\labp'} x : \tau^{\labt' \sqcap \labp'}$\\
and if $\ast \lab_r'$ then $\ast\labp$, then $\ast \labt$, and then $\ast (\labt \sqcap \labp')$\\
and $\labb \sqcap \labp' \sqsubseteq \labt' \sqcap \labp'$.\\
If $\labb \sqsubseteq \labp'$ then $\labb \sqsubseteq \labt' \sqcap \labp'$.\\
$~~~$ By \trule{write} $\Gamma \vdash_{\labp'} \omega := x : \mathbf{Unit}^{\labp'}$.\\
Otherwise $\labp' \sqsubset \labb$, so that $\ast \labb$.\\
$~~~$ Because $\labb \sqsubseteq \labt' \sqsubseteq \labp$, we have $\ast \labp$ and thus $\ast \labt$.\\
$~~~$ By \trule{value} $ \omega :
    \mathbf{Obj}(\tau^{\labb})^{\labt''} \in \Gamma$ and $\labt \sqsubseteq \labt''$.\\
$~~~$ Then $\ast \labt''$.\\
$~~~$ By \trule{write stuck} $\Gamma \vdash_{\labp} \omega := x : \mathbf{Stuck}$. \\
$~~~$ By \trule{subsumption stuck-II} 
$~~~~~~$ $\Gamma \vdash_{\labp} \omega := x : \mathbf{Unit}^{\labp'}$. \\
\item[Case (Typ execute)]
$$\inferrule*[right= ~\rem{\ast\labp \Rightarrow \ast\labt}]
    {\omega : \mathbf{Obj}((\nabla_{\labp''}.~\mathbf{Bin}(\tau^{\labt'}))^{\labb})^{\labt} \in \Gamma \quad
    \labp \sqsubseteq \labp'' \sqcap \labb}
    {\Gamma \vdash_{\labp} \mathsf{exec}~\omega : \tau^{\labt' \sqcap \labp}}
$$
$\labp' \sqsubset \labp \sqsubseteq \labp'' \sqcap \labb$\\
and if $\ast \lab_r'$ then $\ast\labp$, and then $\ast \labt$.\\
By \trule{execute} $\Gamma \vdash_{\labp'} \mathsf{exec}~\omega : \tau^{\labt' \sqcap \labp'}$.\\
Here $\labt' \sqcap \labp' = \labt' \sqcap \labp \sqcap \labp'$.\\
\item[Case (Typ read)]
$$
\inferrule*[right = ~    \rem{\ast(\labb \sqcap \labp)\Rightarrow \ast\labt}]
    {\omega : \mathbf{Obj}(\tau^{\labb})^{\labt} \in \Gamma}
    {\Gamma \vdash_{\labp} ~!\omega : \tau^{\labb \sqcap \labp}}
$$
If $\ast (\labb \sqcap \labp')$ then $\ast(\labb \sqcap \labp)$, and then $\ast \labt$.\\
By \trule{read} $\Gamma \vdash_{\labp'} ~!\omega : \tau^{\labb \sqcap \labp'}$.\\
Here $\labb \sqcap \labp' = \labb \sqcap \labp \sqcap \labp'$.\\
\item[Case (Typ limit)]
$$\infer
    {\Gamma
    \vdash_{\labp''} a : T}
    {\Gamma \vdash_{\labp} [\labp'']~a : T}
$$
Let $T = \tau^{\labt}$.\\
Then $\labt \sqsubseteq \labp''$.\\
If $\labp'' \sqsubseteq \labp'$ then\\
$~~~$ $\labt \sqcap \labp' = \labt$.\\
$~~~$ By \trule{limit} $\Gamma \vdash_{\labp'} [\labp'']~a : \tau^{\labt \sqcap \labp'}$.\\
Otherwise $\labp' \sqsubset \labp''$.\\
By \trule{escalate stuck} $\Gamma \vdash_{\labp'} [\labp'']~a : \mathbf{Stuck}$.\\
By \trule{subsumption stuck-II} $\Gamma \vdash_{\labp'} [\labp'']~a : \tau^{\labt \sqcap \labp'}$.\\
\item[Case (Typ evaluate)]
$$\infer
    {\Gamma \vdash_{\labp} a : T' \\
    \Gamma,x: T' \vdash_{\labp} b : T}
    {\Gamma \vdash_{\labp} \eval x a b : T}
$$
Let $T = \tau^{\labt}$.\\
By I.H. $\Gamma \vdash_{\labp'} a : T''$ and $\Gamma, x : T'' \vdash_{\labp'} b : \tau^{\labt \sqcap \labp'}$.\\
By \trule{evaluate} $\Gamma \vdash_{\labp'} \eval x a b :  \tau^{\labt \sqcap \labp'}$.\\
\item[Case (Typ substitute)]
$$\infer
    {\Gamma \vdash_{\labp'} \mu : T' \\
    ~~~\Gamma, x: T' \vdash_{\labp} a : T}
    {\Gamma \vdash_{\labp} \new{x/\mu @\labp'} a : T}
$$
Let $T = \tau^{\labt}$.\\
By I.H. $\Gamma, x : T' \vdash_{\labp'} a : \tau^{\labt \sqcap \labp'}$.\\
By \trule{substitute} $\Gamma\! \vdash_{\labp'}\! \new{x/\mu @ \labp'}a :  \tau^{\labt \sqcap \labp'}.\!\!\!$\qedhere
\end{description}
\end{proof}
%
\begin{lemma}[Bind]\label{application} Suppose that $a = a'\{x/y\}$. Then $\Gamma \vdash_{\labp} a : \_$ if and only if $\Gamma \vdash_{\labp} \new{x/y@\labp}~a'$.
\end{lemma}
\begin{proof} By induction on the structure of $a'$.
\end{proof}
$ $\\
{\bf Restatement of Theorem \ref{subjred}} (Type preservation) {\em
  Suppose that $\Gamma \vdash_{\labp} \sigma$ and $\Gamma \vdash_{\labp} a : \_$. Then 
\begin{enumerate}
\item If $a \equiv b$ then $\Gamma \vdash_{\labp} b
  : \_$.
\item If $a \action{\labp;\sigma} b$ then $\Gamma \vdash_{\labp} b : \_$.
\end{enumerate}
}
\begin{proof}[Proof of (1)] We prove preservation under $\equiv$ by induction on the structure of derivations.
\begin{defn2}
\mycategory{\mathcal E_{\lab;\sigma}}{evaluation context} \\
\entry{\bullet_{\lab;\sigma}}{hole} \\
\entry{\eval x {\mathcal E_{\lab;\sigma}} b}{evaluate sequential} \\
\entry{\fork{\mathcal E_{\lab;\sigma}} b}{fork child} \\
\entry{\fork a \mathcal E_{\lab;\sigma}}{fork parent} \\
\entry{\new {x/\mu @\lab'} \mathcal E_{\lab;\{x/\mu@\lab'\}\cup \sigma}}{restrict substitution} \\
\entry{[\lab']~\mathcal E_{\lab';\sigma}~~~~~(\lab' \sqsubseteq \lab)}{lower process label}
\end{defn2}
$ $\\
{\bf Case (Struct substitution)}
$$\infer
    {x \notin \fv(\mathcal E_{\lab;\sigma}) \cup \bv(\mathcal E_{\lab;\sigma}) \\ \fv(\mu) \cap \bv(\mathcal E_{\lab;\sigma}) = \varnothing}
    {\new {x/\mu@\lab''} \llctx{\lab,\{x/\mu@\lab''\}\cup \sigma} a {\lab';\sigma'} \equiv \sctx{\new {x/\mu@\lab''} a} {\lab';\sigma'}}
$$
$~~~$ Let $\sigma'' = \{x/\mu@\lab''\}\cup \sigma$.
\begin{itemize}
\item 
$\new{x/\mu@\lab''}\eval y {\llctx{\lab;\sigma''}{a'}{\lab';\sigma'}}{b'}$\\
$~~~~~~~ \equiv\eval y {\sctx{\new{x/\mu@\lab''}a'}{\lab';\sigma'}}{b'}$\\
and $\Gamma' \vdash_{\lab} \new{x/\mu@\lab''}\eval y{~\llctx{\lab;\sigma''}{a'}{\lab';\sigma'}}{b'}
  : T$.\\
  \\
By \trule{substitute} and \trule{evaluate}\\
$~~~$ $\Gamma' \vdash_{\lab''} \mu
  : T''$\\
$~~~$ and $\Gamma', x: T'' \vdash_{\lab} \llctx{\lab;\sigma''}{a'}{\lab';\sigma'}
  : T'''$\\
$~~~$ and $\Gamma', x: T'', y : T''' \vdash_{\lab} b'
  : T$.\\
By \trule{substitute} and S.R.\\
$~~~$ $\Gamma' \vdash_{\lab} \new{x/\mu@\lab''}\llctx{\lab;\sigma''}{a'}{\lab';\sigma'}
  : T'''$\\
$~~~$ and $\Gamma', y : T''' \vdash_{\lab} b'
  : T$.\\
By I.H. $\Gamma' \vdash_{\lab} \sctx{\new{x/\mu@\lab''}a'}{\lab';\sigma'}
  : T'''$.\\
By \trule{evaluate}\\
$~~~$ $\Gamma' \vdash_{\lab} \eval y {\sctx{\new{x/\mu@\lab''}a'}{\lab';\sigma'}}{b'}
  : T$.\\
\item 
$\new{x/\mu@\lab''}\fork {\llctx{\lab;\sigma''}{a'}{\lab';\sigma'}}{b'}$\\
$~~~~~~~ \equiv\fork {\sctx{\new{x/\mu@\lab''}a'}{\lab';\sigma'}}{b'}$\\
$~~~$ and $\Gamma' \vdash_{\lab} \new{x/\mu@\lab''}\fork {\llctx{\lab;\sigma''}{a'}{\lab';\sigma'}}{b'}
  : T$.\\
  \\
By \trule{substitute} and \trule{fork} \\
$~~~$ $\Gamma' \vdash_{\lab''} \mu
  : T''$\\
$~~~$ and $\Gamma', x: T'' \vdash_{\lab} \llctx{\lab;\sigma''}{a'}{\lab';\sigma'}
  : T'''$\\
$~~~$ and $\Gamma', x: T'' \vdash_{\lab} b'
  : T$.\\
By \trule{substitute} and S.R.\\
$~~~$ $\Gamma' \vdash_{\lab} \new{x/\mu@\lab''}\llctx{\lab;\sigma''}{a'}{\lab';\sigma'}
  : T'''$\\
$~~~$ and $\Gamma' \vdash_{\lab} b'
  : T$.\\
By I.H. $\Gamma' \vdash_{\lab} \sctx{\new{x/\mu@\lab''}a'}{\lab';\sigma'}
  : T'''$.\\
By \trule{fork}\\
$~~~$ $\Gamma' \vdash_{\lab} \fork {\sctx{\new{x/\mu@\lab''}a'}{\lab';\sigma'}}{b'}
  : T$.\\
\item 
$\new{x/\mu@\lab''}\fork{b'} {\llctx{\lab;\sigma''}{a'}{\lab';\sigma'}}$\\
$~~~~~~~ \equiv\fork{b'} {\sctx{\new{x/\mu@\lab''}a'}{\lab';\sigma'}}$\\
$~~~$ and $\Gamma' \vdash_{\lab} \new{x/\mu@\lab''}\fork{b'} {\llctx{\lab;\sigma''}{a'}{\lab';\sigma'}}
  : T$.\\
  \\
By \trule{substitute} and \trule{fork} \\
$~~~$ $\Gamma' \vdash_{\lab''} \mu
  : T''$\\
$~~~$ and $\Gamma', x: T'' \vdash_{\lab} \llctx{\lab;\sigma''}{a'}{\lab';\sigma'}
  : T$\\
$~~~$ and $\Gamma', x: T'' \vdash_{\lab} b'
  : T'''$.\\
By \trule{substitute} and S.R.\\
$~~~$ $\Gamma' \vdash_{\lab} \new{x/\mu@\lab''}\llctx{\lab;\sigma''}{a'}{\lab';\sigma'}
  : T$\\
$~~~$ and $\Gamma' \vdash_{\lab} b'
  : T'''$.\\
By I.H. $\Gamma' \vdash_{\lab} \sctx{\new{x/\mu@\lab''}a'}{\lab';\sigma'}
  : T$.\\
By \trule{fork}\\
$~~~$ $\Gamma' \vdash_{\lab} \fork{b'} {\sctx{\new{x/\mu@\lab''}a'}{\lab';\sigma'}}
  : T$.\\
  %
\item 
$\new{x/\mu@\lab''}\new{y/\mu'@\lab'''} {\llctx{\lab;\sigma''}{a'}{\lab';\sigma'}}$\\
$~~~~~~~ \equiv\new{y/\mu'@\lab'''} {\sctx{\new{x/\mu@\lab''}a'}{\lab';\sigma'}}$\\
$~~~$ and $\Gamma' \vdash_{\lab} \new{x/\mu@\lab''}\new{y/\mu'@\lab'''} {\llctx{\lab;\sigma''}{a'}{\lab';\sigma'}}
  :~T$.\\
  \\
By \trule{substitute} and \trule{substitute} \\
$~~~$ $\Gamma' \vdash_{\lab''} \mu
  : T''$\\
$~~~$ and $\Gamma', x: T'' \vdash_{\lab'''} v
  : T'''$\\
$~~~$ and $\Gamma', x: T'', y : T''' \vdash_{\lab} \llctx{\lab;\sigma''}{a'}{\lab';\sigma'}
  : T$.\\
By \trule{substitute} and S.R.\\
$~~~$ $\Gamma', , y: T''' \vdash_{\lab''} u
  : T''$\\
$~~~$ and $\Gamma' \vdash_{\lab'''} \mu'
  : T'''$\\
$~~~$ and $\Gamma', y : T''',x: T'' \vdash_{\lab} \llctx{\lab;\sigma''}{a'}{\lab';\sigma'}
  : T$.\\
By \trule{substitute} \\
$~~~$ $\Gamma', y : T''' \vdash_{\lab} \new{x/\mu@\lab''}\llctx{\lab;\sigma''}{a'}{\lab';\sigma'}
  : T$.\\
By I.H. $\Gamma', y : T''' \vdash_{\lab} \sctx{\new{x/\mu@\lab''}a'}{\lab';\sigma'}
  : T$.\\
By \trule{substitute} \\
$~~~$ $\Gamma' \vdash_{\lab} \new{y/\mu'@\lab'''} {\sctx{\new{x/\mu@\lab''}a'}{\lab';\sigma'}}
  : T$.\\
\item 
$\new{x/\mu@\lab''}[\lab''']~\llctx{\lab''';\sigma''}{a'}{\lab';\sigma'}$\\
$~~~~~~~ \equiv[\lab''']~{\llctx{\lab''';\sigma}{\new{x/\mu@\lab''}a'}{\lab';\sigma'}}$\\
$~~~$ and $\Gamma' \vdash_{\lab} \new{x/\mu@\lab''}[\lab''']~{\llctx{\lab''';\sigma''}{a'}{\lab';\sigma'}}
  : T$.\\
  \\
By \trule{substitute} and \trule{limit} \\
$~~~$ $\Gamma' \vdash_{\lab''} \mu
  : T''$ \\
$~~~$ and $\Gamma', x: T'' \vdash_{\lab'''} \llctx{\lab''';\sigma''}{a'}{\lab';\sigma'}
  : T$.\\
By \trule{substitute} \\
$~~~$ $\Gamma' \vdash_{\lab'''} \new{x/\mu@\lab''}\llctx{\lab''';\sigma}{a'}{\lab';\sigma'}
  : T$.\\
By I.H. $\Gamma' \vdash_{\lab'''} \llctx{\lab''';\sigma}{\new{x/\mu@\lab''}a'}{\lab';\sigma'}
  : T$.\\
By \trule{limit} \\
$~~~$ $\Gamma' \vdash_{\lab} [\lab''']~{\llctx{\lab''';\sigma}{\new{x/\mu@\lab''}a'}{\lab';\sigma'}}
  : T$.
\end{itemize}
{\bf Case (Struct fork)}
$$\infer
    {\fv(a) \cap \bv(\mathcal E_{\lab;\sigma}) = \emptyset}
    {\fork a \sctx b \lab\equiv \sctx{\fork a b} \lab}
$$
\begin{itemize}
\item 
$\fork{a''}\eval x {\sctx{a'}{\lab}}{b'}$\\
$~~~~~~~ \equiv\eval x {\sctx{\fork{a''}a'}{\lab}}{b'}$\\
$~~~$ and $\Gamma' \vdash_{\lab} \fork{a''}\eval x {\sctx{a'}{\lab}}{b'}
  : T$.\\
  \\
By \trule{fork} and \trule{evaluate} \\
$~~~$ $\Gamma' \vdash_{\lab} a''
  : T''$\\
$~~~$ and $\Gamma' \vdash_{\lab} \sctx{a'}{\lab}
  : T'''$\\  
$~~~$ and $\Gamma', x : T''' \vdash_{\lab} b'
  : T$.\\
By \trule{fork} \\
$~~~$ $\Gamma' \vdash_{\lab} \fork{a''}{\sctx{a'}{\lab}}
  : T'''$\\  
$~~~$ and $\Gamma', x : T''' \vdash_{\lab} b'
  : T$.\\
By I.H. $\Gamma' \vdash_{\lab} \sctx{\fork{a''}a'}{\lab}
  : T'''$.\\
By \trule{evaluate} \\
$~~~$ $\Gamma' \vdash_{\lab} \eval x {\sctx{\fork{a''}a'}{\lab}}{b'}
  : T$.\\
\item 
$\fork{a''}\fork{\sctx{a'}{\lab}}{b'}$\\
$~~~~~~~ \equiv\fork{\sctx{\fork{a''}a'}{\lab}}{b'}$\\
$~~~$ and $\Gamma' \vdash_{\lab} \fork{a''}\fork{\sctx{a'}{\lab}}{b'}
  : T$.\\
  \\
By \trule{fork} and \trule{fork} \\
$~~~$ $\Gamma' \vdash_{\lab} a''
  : T''$\\
$~~~$ and $\Gamma' \vdash_{\lab} \sctx{a'}{\lab}
  : T'''$\\  
$~~~$ and $\Gamma' \vdash_{\lab} b'
  : T$.\\
By \trule{fork} \\
$~~~$ $\Gamma' \vdash_{\lab} \fork{a''}{\sctx{a'}{\lab}}
  : T'''$\\  
$~~~$ and $\Gamma' \vdash_{\lab} b'
  : T$.\\
By I.H. $\Gamma' \vdash_{\lab} \sctx{\fork{a''}a'}{\lab}
  : T'''$.\\
By \trule{fork} \\
$~~~$ $\Gamma' \vdash_{\lab} \fork{\sctx{\fork{a''}a'}{\lab}}{b'}
  : T$.\\
\item 
$\fork{a''}\fork{b'}{\sctx{a'}{\lab}}$\\
$~~~~~~~ \equiv\fork{b'}{\sctx{\fork{a''}a'}{\lab}}$\\
$~~~$ and $\Gamma' \vdash_{\lab} \fork{a''}\fork{b'}{\sctx{a'}{\lab}}
  : T$.\\
  \\
By \trule{fork} and \trule{fork} \\
$~~~$ $\Gamma' \vdash_{\lab} a''
  : T''$\\
$~~~$ and $\Gamma' \vdash_{\lab} b'
  : T'''$\\  
$~~~$ and $\Gamma' \vdash_{\lab} \sctx{a'}{\lab}
  : T$.\\
By \trule{fork} \\
$~~~$ $\Gamma' \vdash_{\lab} b'
  : T'''$\\  
$~~~$ and $\Gamma' \vdash_{\lab} \fork{a''}{\sctx{a'}{\lab}}
  : T$.\\
By I.H. $\Gamma' \vdash_{\lab} \sctx{\fork{a''}a'}{\lab}
  : T$.\\
By \trule{fork} \\
$~~~$ $\Gamma' \vdash_{\lab} \fork{b'}{\sctx{\fork{a''}a'}{\lab}}
  : T$.\\
  %
\item 
$\fork{a''}\new{x/\mu@\lab'}{\sctx{a'}{\lab}}$\\
$~~~~~~~ \equiv\new{x/u@\lab'}{\sctx{\fork{a''}a'}{\lab}}$\\
$~~~$ and $\Gamma' \vdash_{\lab} \fork{a''}\new{x/u@\lab'}{\sctx{a'}{\lab}}
  : T$.\\
  \\
By \trule{fork} and \trule{substitute} \\
$~~~$ $\Gamma' \vdash_{\lab} a''
  : T''$\\
$~~~$ and $\Gamma' \vdash_{\lab';\sigma'} \mu
  : T$\\
$~~~$ and $\Gamma', x: T''' \vdash_{\lab} \sctx{a'}{\lab}
  : T$.\\
By S.R. $\Gamma', x: T'''  \vdash_{\lab} a''
  : T''$.\\
By \trule{fork} \\
$~~~$ $\Gamma', x: T''' \vdash_{\lab} \fork{a''}\sctx{a'}{\lab}
  : T$.\\
By I.H. $\Gamma', x: T''' \vdash_{\lab} \sctx{\fork{a''}a'}{\lab}
  : T$.\\
By \trule{substitute} \\
$~~~$ $\Gamma' \vdash_{\lab} \new{x/u@\lab'}{\sctx{\fork{a''}a'}{\lab}}
  : T$.
\end{itemize}
{\bf Case (Struct store)}
    $$\infer
    {}
    {\fork{\omega \store\lab u}[\lab']~a \equiv [\lab']~(\fork{\omega \store\lab u} a)}
    $$
$\fork{\omega \store{\lab''} u}[\lab']~a'$\\
$~~~~~~~ \equiv [\lab']~(\fork{\omega \store{\lab''} u} a')$\\
$~~~$ and $\Gamma' \vdash_{\lab} \fork{\omega \store{\lab''} u}[\lab']~a'
  : T$.\\
  \\
By \trule{fork} \\
$~~~$ $\Gamma' \vdash_{\lab} \omega \store{\lab''} u
  : \_$ \\
$~~~$ and $\Gamma' \vdash_{\lab} [\lab']~a'
  : T$.\\
By \trule{limit} \\
$~~~$ $\Gamma' \vdash_{\lab'} \omega \store{\lab''} u
  : \_$ \\
$~~~$ and $\Gamma' \vdash_{\lab'} a'
  : T$.\\
By \trule{fork} $\Gamma' \vdash_{\lab'} \fork{\omega \store{\lab''} u}a'
  : T$.\\    
By \trule{limit} $\Gamma' \vdash_{\lab} [\lab']~\fork{\omega \store{\lab''} u}a'
  : T$.\\
\\
%
%
{\bf Case (Struct bind)}\\
By Lemma \ref{application}.
\end{proof}
\begin{proof}[Proof of (2)] We prove preservation under $\longrightarrow$ by induction on the structure of derivations.\\
\\
{\bf Case (Reduct evaluate)}
$$\infer
    {}
    {\eval x u a \action{\lab;\sigma} \new{x/u@\lab} a}
$$
$\Gamma \vdash_\lab \eval x u a' : T$.\\
\\
By \trule{evaluate} \\
$~~~$ $\Gamma \vdash_\lab u : T''$ \\
$~~~$ and $\Gamma, x: T'' \vdash_\lab a' : T$.\\
By \trule{substitute}  $\Gamma \vdash_\lab \new{x/u@\lab} a' : T$.\\
\\
{\bf Case (Reduct new)}
$$\infer
    {}
    {\mathsf{new}(x\mbox{ \# }\labb) \action{\labp;\sigma} \new{\omega/\mathsf{new}(x\mbox{ \# }\labb)@\labp} (\omega \store{\labp} x \Rsh \omega)}
$$
$\Gamma \vdash_{\labp} \mathsf{new}(x\mbox{ \# }\labb) : T$.\\
\\
By \trule{new} \\
$~~~$ $\Gamma \vdash_{\labp} x : \tau^{\labt}$, \\
$~~~$ $\labb \sqsubseteq \labt$, \\
$~~~$ and $T = \mathbf{Obj}(\tau^{\labb})^{\labp}$.\\
By \trule{store} $\Gamma, \omega: T \vdash_{\labp} \omega\store{\labp} x : \_$.\\
By \trule{fork} $\Gamma, \omega:T \vdash_{\labp} \omega\store{\labp} x \Rsh \omega : T$.\\
By \trule{substitute} \\ 
$~~~$ $\Gamma \vdash_{\labp}  \new{\omega/\mathsf{new}(x\mbox{ \# }\labb)@\labp} (\omega \store{\labp} x \Rsh \omega) : T$.\\
\\
{\bf Case (Reduct read)}
$$\infer
    {\omega \stackrel\sigma= \omega'}
    {\fork{\omega \store\lab x} ~!\omega' \action{\lab';\sigma} \fork{\omega \store\lab x}x}
$$
$\Gamma \vdash_\lab \fork{\omega \store{\labo} x} ~!\omega' : \tau^{\labt}$.\\
\\
By \trule{fork}\\
$~~~$ $\Gamma \vdash_\lab \omega \store{\labo} x : \_$. \\
By \trule{store} $\Gamma \vdash_\lab x : \_$.\\
By \trule{fork} $\Gamma \vdash_\lab \omega \store{\labo} x \Rsh x : \_$. \\
\\
{\bf Case (Reduct write)}
$$\infer
    {\omega \stackrel\sigma=\omega' \\ \lab \sqsubseteq \lab'}
    {\fork{\omega \store\lab x} \omega' := x' \action{\lab';\sigma} \fork{\omega \store\lab x'} \mathsf{unit}}
$$
$\Gamma \vdash_\lab \fork{\omega \store{\labo} x} \omega' := x' : \mathbf{Unit}^\lab$.\\
\\
By \trule{fork}\\
$~~~$ $\Gamma \vdash_\lab \omega \store{\labo} x : \_$ \\
$~~~$ and $\Gamma \vdash_\lab \omega' := x' : \mathbf{Unit}^{\lab}$\\
and $\labo \sqsubseteq \lab$.\\
\\
By \trule{store}, \trule{write}, and $\Gamma \vdash \sigma$\\
$~~~$ $\omega: \mathbf{Obj}(\tau^{\labb})^\_ \in \Gamma$,\\
$~~~$ $\labb \sqsubseteq \labo$, \\
$~~~$ $\Gamma \vdash_\lab \omega' : \mathbf{Obj}(\tau^{\labb})^{\labt}$, \\
$~~~$ $\Gamma \vdash_\lab x' : \tau^{\labt'}$, \\
$~~~$ and $\labb \sqsubseteq \labt'$.\\
\\
By \trule{store} $\Gamma \vdash_\lab \omega \store{\labo} x' : \_$. \\
By \trule{unit} $\Gamma \vdash_\lab \omega \store{\labo} \mathsf{unit} : \mathbf{Unit}^\lab$. \\
By \trule{fork} $\Gamma \vdash_\lab \omega \store{\labo} x' \Rsh \mathsf{unit} : \mathbf{Unit}^\lab$.\\
\\
{\bf Case (Reduct execute)}
$$\infer
    {\omega \stackrel\sigma=\omega' \\ \mathsf{pack}(f) \in \sigma(x) \\ \lab'' = \lab' \sqcap \lab}
    {\fork{\omega \store\lab x}\mathsf{exec}~\omega' \action{\lab';\sigma} \fork{\omega \store\lab x}[\lab'']~f}
$$
$\Gamma \vdash_\lab \fork{\omega \store{\labo} x} \mathsf{exec}~\omega' : \_$.\\
\\
By \trule{fork}\\
$~~~$ $\Gamma \vdash_\lab \omega \store{\labo} x : \_$ \\
$~~~$ and $\Gamma \vdash_\lab \mathsf{exec}~\omega' : \_$.\\
\\
By \trule{store}, \trule{execute}, and $\Gamma \vdash \sigma$\\
$~~~$ $\Gamma \vdash_{\labp'} \mathsf{pack}(f): \nabla_{\labp}.~\mathbf{Bin}(T)^{\labp'}$ for some $\labp'$,\\
$~~~$ $x : \nabla_{\labp}.~\mathbf{Bin}(T)^{\labt} \in \Gamma$,\\
$~~~$ $\omega: \mathbf{Obj}(\nabla_{\labp}.~\mathbf{Bin}(T)^{\labb})^\_ \in \Gamma$,\\
$~~~$ $\labb \sqsubseteq \labo \sqcap \labt$,\\
$~~~$ and $\lab \sqsubseteq \labp \sqcap \labb$. \\
\\
By \trule{pack} $\Gamma \vdash_{\labp} f: \_$.\\
By \trule{subsumption process label} $\Gamma \vdash_{\lab} f: \_$.\\
By \trule{fork} $\Gamma \vdash_{\lab} \omega \store{\labo} x  \Rsh f: \_$.\\
\\
{\bf Case (Reduct un/protect)}
$$\infer
    {\omega \stackrel\sigma=\omega' \\ \lab \sqcup \lab' \sqsubseteq \lab''}
    {\fork{\omega \store{\lab} x}\langle\lab'\rangle~\omega' \action{\lab'';\sigma} \fork{\omega \store{\lab'} x}\mathsf{unit}}
$$
$\Gamma \vdash_\lab \fork{\omega \store{\labo} x} \langle\lab'\rangle~\omega' : \mathbf{Unit}^\lab$.\\
\\
By \trule{fork}\\
$~~~$  $\Gamma \vdash_\lab \omega \store{\labo} x : \_$ \\
$~~~$ and $\Gamma \vdash_\lab \langle\lab'\rangle~\omega' : \mathbf{Unit}^\lab$\\
$~~~$ $\labo \sqcup \lab' \sqsubseteq \lab$.\\
\\
By \trule{store}, \trule{un/protect}, and $\Gamma \vdash \sigma$, \\
$~~~$ $\omega: \mathbf{Obj}(\tau^{\labb})^\_ \in \Gamma$,\\
$~~~$ $\labb \sqsubseteq \labo$, \\
$~~~$ $\Gamma \vdash_\lab \omega': \mathbf{Obj}(\tau^{\labb})^\_$,\\
$~~~$ and $\labb \sqsubseteq \lab'$. \\
\\
By \trule{store} $\Gamma \vdash_\lab \omega \store{\lab'} x : \_$.\\
By \trule{unit} $\Gamma \vdash_\lab \mathsf{unit} : \mathbf{Unit}^\lab$.\\
By \trule{fork} $\Gamma \vdash_\lab \fork{\omega \store{\lab'} x} \mathsf{unit} : \mathbf{Unit}^\lab$.\\
\\
{\bf Case (Reduct context)}
$$\infer
    {a \action{\lab';\sigma'} b}
    {\sctx a {\lab';\sigma'}\action{\lab;\sigma} \sctx b {\lab';\sigma'}}
$$
\begin{defn2}
\mycategory{\mathcal E_{\lab;\sigma}}{evaluation context} \\
\entry{\bullet_{\lab;\sigma}}{hole} \\
\entry{\eval x {\mathcal E_{\lab;\sigma}} b}{evaluate sequential} \\
\entry{\fork{\mathcal E_{\lab;\sigma}} b}{fork child} \\
\entry{\fork a \mathcal E_{\lab;\sigma}}{fork parent} \\
\entry{\new {x/\mu @\lab'} \mathcal E_{\lab,\{x/\mu@\lab'\}\cup \sigma}}{restrict substitution} \\
\entry{[\lab']~\mathcal E_{\lab';\sigma}~~~~~(\lab' \sqsubseteq \lab)}{lower process label}
\end{defn2}
\begin{itemize}
\item $\eval x {\sctx{a'}{\lab';\sigma'}} b' \action{\lab;\sigma} \eval x {\sctx{a''}{\lab';\sigma'}} b'$,\\
$~~~$ $a' \action{\lab';\sigma'} a''$,\\
$~~~$ and $\Gamma \vdash_\lab \eval x {\sctx{a'}{\lab';\sigma'}} b' : T$.\\
\\
By (\rrule{context}) and \trule{evaluate} \\
$~~~$ $\sctx{a'}{\lab';\sigma'} \action{\lab;\sigma} \lctx{a''}{\lab';\sigma'}$,\\
$~~~$ $\Gamma \vdash_\lab \sctx{a'}{\lab';\sigma'} : T''$, \\
$~~~$ and $\Gamma, x: T'' \vdash_\lab b' : T$.\\
By I.H. $\Gamma \vdash_\lab \sctx{a''}{\lab';\sigma'} : T''$.\\
By \trule{evaluate} \\
$~~~$ $\Gamma \vdash_\lab \eval x {\sctx{a''}{\lab';\sigma'}} b' : T$.\\
\item $\fork {\sctx{a'}{\lab';\sigma'}} b' \action{\lab;\sigma} \fork {\sctx{a''}{\lab';\sigma'}} b'$,\\
$~~~$ $a' \action{\lab';\sigma'} a''$,\\
$~~~$ and $\Gamma \vdash_\lab \fork {\sctx{a'}{\lab';\sigma'}} b' : T$.\\
\\
By (\rrule{context}) and \trule{fork} \\
$~~~$ $\sctx{a'}{\lab';\sigma'} \action{\lab;\sigma} \sctx{a''}{\lab';\sigma'}$,\\
$~~~$ $\Gamma \vdash_\lab \sctx{a'}{\lab';\sigma'} : T''$, \\
$~~~$ and $\Gamma \vdash_\lab b' : T$.\\
By I.H. $\Gamma \vdash_\lab \sctx{a''}{\lab';\sigma'} : T''$.\\
By \trule{fork} \\
$~~~$ $\Gamma \vdash_\lab \fork {\sctx{a''}{\lab';\sigma'}} b' : T$.\\
\item $\fork {b'}{\sctx{a'}{\lab';\sigma'}} \action{\lab;\sigma} \fork {b'} {\sctx{a''}{\lab';\sigma'}}$,\\
$~~~$ $a' \action{\lab';\sigma'} a''$,\\
$~~~$ and $\Gamma \vdash_\lab \fork {b'}{\sctx{a'}{\lab';\sigma'}} : T$.\\
\\
By (\rrule{context}) and \trule{fork} \\
$~~~$ $\sctx{a'}{\lab';\sigma'} \action{\lab;\sigma} \sctx{a''}{\lab';\sigma'}$,\\
$~~~$ $\Gamma \vdash_\lab \sctx{a'}{\lab';\sigma'} : T$, \\
$~~~$ and $\Gamma \vdash_\lab b' : T''$.\\
By I.H. $\Gamma \vdash_\lab \sctx{a''}{\lab';\sigma'} : T$.\\
By \trule{fork} \\
$~~~$ $\Gamma \vdash_\lab \fork {b'}{\sctx{a''}{\lab';\sigma'}} : T$.\\
%
\item $\new{x/u@\lab''}{\sctx{a'}{\lab';\sigma'}} \action{\lab;\sigma} \new{x/u@\lab''}{\sctx{a''}{\lab';\sigma'}}$,\\
$~~~$ $a' \action{\lab';\sigma'} a''$,\\
$~~~$ and $\Gamma \vdash_\lab \new{x/u@\lab''}{\sctx{a'}{\lab';\sigma'}} : T$.\\
\\
By (\rrule{context}) and \trule{substitute} \\
$~~~$ $\lctx{a'}{\lab';\sigma'} \action{\lab;\sigma} \sctx{a''}{\lab';\sigma'}$,\\
$~~~$ and $\Gamma \vdash_{\lab''} u : T''$,\\
$~~~$ and $\Gamma, x : T'' \vdash_\lab \sctx{a'}{\lab';\sigma'} : T$.\\
By I.H. $\Gamma, x : T'' \vdash_\lab \sctx{a''}{\lab';\sigma'} : T$.\\
By \trule{substitute} \\
$~~~$ $\Gamma \vdash_\lab \new{x/u@\lab''}{\sctx{a''}{\lab';\sigma'}} : T$.\\
\item $[\lab'']~{\llctx{\lab'';\sigma}{a'}{\lab';\sigma'}} \action{\lab;\sigma} [\lab'']~{\llctx{\lab'';\sigma}{a''}{\lab';\sigma'}}$,\\
$~~~$ $a' \action{\lab';\sigma'} a''$,\\
$~~~$ and $\Gamma \vdash_\lab [\lab'']~{\llctx{\lab'';\sigma}{a'}{\lab';\sigma'}} : T$.\\
\\
By (\rrule{context}) and \trule{limit} \\
$~~~$ $\llctx{\lab'';\sigma}{a'}{\lab';\sigma'} \action{\lab'';\sigma} \llctx{\lab'';\sigma}{a''}{\lab';\sigma'}$\\
$~~~$ and $\Gamma \vdash_{\lab''} \llctx{\lab'';\sigma}{a'}{\lab';\sigma'} : T$.\\
By I.H. $\Gamma \vdash_{\lab''} \llctx{\lab'';\sigma}{a''}{\lab';\sigma'} : T$.\\
By \trule{limit} \\
$~~~$ $\Gamma \vdash_\lab [\lab'']~{\llctx{\lab'';\sigma}{a''}{\lab';\sigma'}} : T$.
\end{itemize}
{\bf Case (Reduct congruence)}
$$\infer
    {a \equiv a' \\ a' \action{\lab;\sigma} b' \\ b' \equiv b}
    {a \action{\lab;\sigma} b}
$$
$\Gamma \vdash_\lab a : T$,\\
$~~~$ $a \equiv a'$,\\
$~~~$ $a' \action{\lab;\sigma} b'$, \\ 
$~~~$ and $b' \equiv b$.\\
\\
By Theorem \ref{subjred}(1) $\Gamma \vdash_\lab a' : \_$.\\
By I.H. $\Gamma \vdash_\lab b' : \_$.\\
So by Theorem \ref{subjred}(1) $\Gamma \vdash_\lab b : \_$.
\end{proof}
$ $\\
{\bf Restatement of Theorem \ref{mainthm}} (Enforcement of strong DFI) {\em Let
  $\Omega$ be the set of objects whose contents are trusted beyond $\lab$ in $\Gamma$. Suppose that $\Gamma
  \vdash_\top a : \_~\mathsf{despite}~\labc$, where $\labc \sqsubseteq \lab$. Then $a$
  protects $\Omega$ from $\lab$ despite $\labc$. 
}
%
\begin{proof} Let $e$ be any $\labc$-adversary $[\labc]~e'$. \\
By Proposition \ref{advtyp} $\Gamma \vdash_{\top} e : \_$. \\
By \trule{fork} $\Gamma \vdash_\top \fork a e :_{\_} \_$. \\
\\
Suppose that $\omega \in \Omega$. We need to prove that there are no $\sigma$ and $x$ such that $\fork a [\labc]~e'\action{\top}\!\!\!\!{}^\star~ \llctx{\top;\varnothing}{\omega\store{\_} x}{\top;\sigma}$
and $x\stackrel\sigma\blacktriangledown \lab$. Assume otherwise.\\
\\
By Theorem \ref{subjred} there exists $\Gamma'$ extending $\Gamma$ such that \\
$~~~\Gamma' \vdash \sigma$ and $\Gamma' \vdash_\top \omega \store\_ x : \_$.\\
By \trule{store} $\omega : \mathbf{Obj}(\tau^{\labb})^\_ \in \Gamma'$ such that $\labb \sqsubseteq \labt$.\\
We proceed by induction on the derivation of $x\stackrel\sigma\blacktriangledown \lab$.
\begin{description}
\item[Case] $\labp \sqsubseteq \lab$.\\
For some $\tau$ and $\labt$, $\Gamma' \vdash_{\labp} \mu : \tau^{\labt}$.\\
Then $\labt \sqsubseteq \labp$ and by \trule{value} $\Gamma' \vdash_{\top} x : \tau^{\labt}$.\\
Then $\labt \sqsubseteq \lab$.\\
Then $\labb \sqsubseteq \lab$.\\
But by assumptions $\labb \sqsupset \lab$ (contradiction).
\item[Case] $\mu \equiv y$ for some $y$ and $y\stackrel\sigma\blacktriangledown \lab$.\\
By I.H. $\Gamma' \vdash_{\top} y : \tau^{\labt}$ for some $\labt$ such that $\labt \sqsubseteq \lab$.\\
Then $\labb \sqsubseteq \lab$.\\
But by assumptions $\labb \sqsupset \lab$ (contradiction). \qedhere
\end{description}
\end{proof}
%
%
\noindent
{\bf Restatement of Theorem \ref{optim}} (Redundancy of execution control) {\em Suppose that $\Gamma \vdash_{\top} a : \_ ~\mathsf{despite}~\labc$ and  
$a \action{\top;\varnothing}\!\!\!\!{}^\star ~\llctx{\top;\varnothing}{\fork{\omega \store{\labo} x} \mathsf{exec}\:\omega'}{\labp;\sigma}$ such that $\omega \stackrel\sigma= \omega'$, and $\labp \sqsupset\labc$. Then $\labp \sqsubseteq \labo$.
%
}
\begin{proof} The proof is by inspection of Case (Reduct execute) in the proof of Theorem \ref{subjred}. Recalling that case (where $\lab$ is the process label): $\lab \sqsubseteq \labb \sqsubseteq \labo$.
\end{proof}

\section{An efficient typechecking algorithm}\label{algo}
\begin{figure}
\cenvvv{Typechecking judgments for processes}{\Gamma \vdash_{\labp} a : T \rhd \Gamma'}{
({\bf Typc value}) \vspace{-1mm}
$$\infer
    {x : \tau^{\labt} \in \Gamma}
    {\Gamma \vdash_{\labp} x : \tau^{\labt \sqcap \labp} \rhd \varnothing}
$$
~

({\bf Typc new}) \vspace{-1mm}
$$\infer
    {\Gamma \vdash_{\labp} u : \tau^{\labt} \rhd \varnothing \\ \dots}
    {\Gamma \vdash_{\labp} \mathsf{new}(u\mbox{ \# }\labb) : \mathbf{Obj}(\tau^{\labb})^{\labp} \rhd \varnothing}
$$
~

({\bf Typc pack}) \vspace{-1mm}
$$
\infer
    {\Gamma \vdash f : T \rhd \Gamma'\\ \Gamma' \models \labp' \\ \Box f}
    {\Gamma \vdash_{\labp} \mathsf{pack}(f) : \chi^{\labp} \rhd \Gamma'_{?\mapsto\labp'},\nabla_{\labp'}.~\mathbf{Bin}(T) <: \chi}
$$
%
%
%
~

({\bf Typc fork}) \vspace{-1mm}
$$
%
\infer
    {\Gamma \vdash_{\labp} a : \_ \rhd \Gamma_1 \\
    \Gamma \vdash_{\labp} b : T \rhd \Gamma_2}
    {\Gamma \vdash_{\labp} \fork a b : T \rhd \Gamma_1,\Gamma_2}
$$
~

({\bf Typc evaluate}) \vspace{-1mm}
$$
\infer
    {\Gamma \vdash_{\labp} a : T' \rhd \Gamma_1 \\
    \Gamma,\Gamma_1,x: T' \vdash_{\labp} b : T \rhd \Gamma_2}
    {\Gamma \vdash_{\labp} \eval x a b : T \rhd \Gamma_1,\Gamma_2}
$$
~

({\bf Typc read}) \vspace{-1mm}
$$
\infer
    {\omega : \mathbf{Obj}(\tau^{\labb})^{\labt} \in \Gamma \\ \dots }
    {\Gamma \vdash_{\labp} ~!\omega : \tau^{\labb \sqcap \labp} \rhd \varnothing}
$$
~

({\bf Typc write}) \vspace{-1mm}
$$
\infer
    {\Gamma \vdash_{\labp} \omega :
    \mathbf{Obj}(\tau_1^{\labb})^{\labt} \\
    \Gamma \vdash_{\labp} u : \tau_2^{\labt'} \\
    \dots}
    {\Gamma \vdash_{\labp} \omega := u : \mathbf{Obj}(\tau^{\labb})^{\labt} \rhd \tau_2 <: \tau_1}
$$
~

({\bf Typc execute}) \vspace{-1mm}
$$
\infer
    {\omega : \mathbf{Obj}(\tau_1^{\labb})^{\labt},  \nabla_{\labp'}.~\mathbf{Bin}(\tau^{\labt'}) <: \tau_1 \in \Gamma \\ \dots}
    {\Gamma \vdash_{\labp} \mathsf{exec}~\omega : \tau^{\labt' \sqcap \labp} \rhd \tau_1 <: \nabla_\labp.~\mathbf{Bin}(\tau^{\labp' \sqcap \labp})}
$$
~
$$
\dots 
$$
}
\end{figure}
\begin{figure}
\cenvvv{Typechecking judgments for expressions}{\Gamma \vdash f : T \rhd \Gamma'}{
({\bf Typc ? value}) \vspace{-2mm}
$$\infer
    {x : \tau^{\labt} \in \Gamma}
    {\Gamma \vdash x : \tau^{\labt \sqcap ?} \rhd \varnothing}
$$
~

({\bf Typc ? limit}) \vspace{-1mm}
$$
\infer
    {\Gamma_{?\mapsto \labp'}
    \vdash_{\labp'} a : T \rhd \Gamma' }
    {\Gamma \vdash [\labp']~a : T \rhd \Gamma'}
$$
~

({\bf Typc ? read}) \vspace{-1mm}
$$
\infer
    {\omega : \mathbf{Obj}(\tau^{\labb})^{\labt} \in \Gamma}
    {\Gamma \vdash ~!\omega : \tau^{\labb \sqcap ?} \rhd \dots }
$$
~

({\bf Typc ? write}) \vspace{-1mm}
$$\infer
    {\Gamma \vdash \omega :
    \mathbf{Obj}(\tau_1^{\labb})^{\labt} \\
    \Gamma \vdash u : \tau_2^{\labt'} }
    {\Gamma \vdash\omega := u : \mathbf{Obj}(\tau^{\labb})^{\labt} \rhd \tau_2 <: \tau_1,  \dots }
$$
~

({\bf Typc ? execute}) \vspace{-1mm}
$$
\infer
    {\omega : \mathbf{Obj}(\tau_1^{\labb})^{\labt},  \nabla_{\labp'}.~\mathbf{Bin}(\tau^{\labt'}) <: \tau_1 \in \Gamma}
    {\Gamma \vdash \mathsf{exec}~\omega : \tau^{\labt' \sqcap ?} \rhd \tau_1 <: \nabla_?.~\mathbf{Bin}(\tau^{\labp' \sqcap ?}), \dots }
$$
~

({\bf Typc ? fork}) \vspace{-1mm}
$$\infer
    {\Gamma \vdash a : \_ \rhd \Gamma_1 \\
    \Gamma \vdash b : T \rhd \Gamma_2}
    {\Gamma \vdash \fork f g : T \rhd \Gamma_1,\Gamma_2}
$$
~

({\bf Typc ? evaluate}) \vspace{-1mm}
$$
\infer
    {\Gamma \vdash a : T' \rhd \Gamma_1 \\
    \Gamma,\Gamma_1,x: T' \vdash b : T \rhd \Gamma_2}
    {\Gamma \vdash \eval x f g : T \rhd \Gamma_1,\Gamma_2}
$$
~
$$
\dots \vspace{-1mm}
$$
}
\vspace{-2mm}
\cenvvv{Satisfiability of constraints}{\Gamma \vdash \diamond}{
({\bf Typc $<:$ obj}) \vspace{-2mm}
$$\infer
	{\Gamma \vdash \diamond}
	{\Gamma,\mathbf{Obj}(T) <: \mathbf{Obj}(T) \vdash \diamond}
$$
~

({\bf Typc $<:$ bin}) \vspace{-1mm}
$$
\infer
	{\labp' \sqsubseteq \labp \\ \Gamma,\tau_1 <:\tau_2 \vdash \diamond}
	{\Gamma,\nabla_{\labp}.~\mathbf{Bin}(\tau_1^{\labt}) <: \nabla_{\labp'}.~\mathbf{Bin}(\tau_2^{\labt \sqcap \labp'}) \vdash \diamond}
$$
~

({\bf Typc $<:$ left}) \vspace{-1mm}
$$\infer
	{\Gamma,\chi <: \nabla_{\labp}.~\mathbf{Bin}(\tau_1^{\labt}), \nabla_{\labp}.~\mathbf{Bin}(\tau_1^{\labt}) <: \nabla_{\labp'}.~\mathbf{Bin}(\tau_2^{\labt'}) \vdash \diamond}
	{\Gamma,\chi <: \nabla_{\labp}.~\mathbf{Bin}(\tau_1^{\labt}), \chi <: \nabla_{\labp'}.~\mathbf{Bin}(\tau_2^{\labt'}) \vdash \diamond}
$$
~

({\bf Typc $<:$ right}) \vspace{-1mm}
$$\infer
	{\Gamma,\nabla_{\labp}.~\mathbf{Bin}(\tau_1^{\labt}) <: \nabla_{\labp'}.~\mathbf{Bin}(\tau_2^{\labt'}), \nabla_{\labp'}.~\mathbf{Bin}(\tau_2^{\labt'}) <: \chi \vdash \diamond}
	{\Gamma,\nabla_{\labp}.~\mathbf{Bin}(\tau_1^{\labt}) <: \chi, \nabla_{\labp'}.~\mathbf{Bin}(\tau_2^{\labt'}) <: \chi \vdash \diamond}
$$
~

({\bf Typc $<:$ middle}) \vspace{-1mm}
$$\infer
	{\Gamma, \nabla_{\labp}.~\mathbf{Bin}(\tau_1^{\labt}) <: \chi, 
	\chi <: \nabla_{\labp'}.~\mathbf{Bin}(\tau_2^{\labt'}), \\
	\qquad\qquad\qquad\qquad\nabla_{\labp}.~\mathbf{Bin}(\tau_1^{\labt}) <: \nabla_{\labp'}.~\mathbf{Bin}(\tau_2^{\labt'}) 
	 \vdash \diamond}
	{\Gamma,\nabla_{\labp}.~\mathbf{Bin}(\tau_1^{\labt}) <: \chi, \chi <: \nabla_{\labp'}.~\mathbf{Bin}(\tau_2^{\labt'}) \vdash \diamond}
$$
~
$$
\dots
$$
}
\end{figure}

\noindent
Finally, we outline an efficient algorithm to mechanize typechecking. Broadly, the algorithm builds constraints and then checks whether those constraints are satisfiable. The only complication is due to $\mathsf{pack}$ processes, which require a ``most general" type. We extend the grammar of types with type variables $\chi$, and introduce a distinguished label $?$ denoting an ``unknown" label. We extend the grammar of typing environments with constraints of the form $\tau_1 <: \tau_2$ and label constraints (\emph{i.e.}, boolean formulae over atoms of the form $\lab_1 \sqsubseteq \lab_2$).  Next, we introduce the following typechecking judgments: 
\begin{itemize}
\item $\Gamma \vdash_{\labp} a : T \rhd \Gamma'$, where $\Gamma'$ contains constraints of the form $\tau_1 <: \tau_2$ only (\emph{i.e.}, the label constraint in $\Gamma'$ is true). 
\item $\Gamma \vdash f : T \rhd \Gamma'$, where $\Gamma'$ may contain a label constraint as well as constraints of the form $\tau_1 <: \tau_2$. 
\end{itemize}
We now present some sample typechecking rules, followed by rules that interpret $<:$. Let us first look at the rules for deriving judgments of the form $\Gamma \vdash_{\labp} a : T \rhd \Gamma'$. These rules build constraints of the form $\tau_1 <: \tau_2$ in $\Gamma'$. We elide by dots ($\dots$) label constraints that appear in the original typing rules. Let $\Gamma'_{? \mapsto \lab}$ denote the typing environment obtained from $\Gamma'$ by replacing all occurrences of $?$ with $\lab$. We write $\Gamma' \models \labp$ iff $\labp$ is the highest $\lab$ for which the label constraint in $\Gamma'_{?\mapsto \lab}$ is true. Note that to derive a judgment of this form for a process $\mathsf{pack}(f)$, we need to derive a judgment of the other form for $f$. In fact, the two kinds of judgments are mutually recursive (see below).

Next, we look at the rules for deriving judgments of the form $\Gamma \vdash f : T \rhd \Gamma'$. These rules apply to expressions that are not explicitly under a change of the process label, \emph{e.g.}, expressions obtained by unpacking $\mathsf{pack}$ processes. They build label constraints from those that appear in the original typing rules; the implicit (unknown) process label is replaced by $?$. Predicate $\Box$ ensures that we do not need to consider $\mathsf{pack}$ processes here; further we can assume that all annotations carry the least trusted label. Once we have an expression of the form $[\labp]~a$, we can derive a judgment of the other form for $a$.

Finally, we look at the rules that interpret constraints of the form $\tau_1 <: \tau_2$. Here $<:$ denotes a subtyping relation that is invariant in $\mathbf{Obj}(\_^{\labb})$ and covariant in $\nabla_{\labp}.~\mathbf{Bin}(\_^{\labt})$, and preserves monotonicity. We introduce the judgment $\Gamma \vdash \diamond$ to check satisfiability of such constraints.

We prove that typechecking is sound and complete.
\begin{proposition}
The typing judgment $\Gamma \vdash_{\labp} a : \_$ can be derived if and only if the typechecking judgment $\Gamma \vdash_{\labp} a : T \rhd \Gamma'$ can be derived for some $T$ and $\Gamma'$ such that $\Gamma' \vdash \diamond$.  
\end{proposition}
\noindent
Further, typechecking terminates in time $\mathcal O(\mathbb L |a|))$ if $\Gamma$ and $a$ have $\mathbb L$ distinct labels. Indeed, let $\varphi(|a|)$ be the running time of the judgment $\Gamma \vdash_{\labp} a : \_ \rhd \_$, and $\psi(|f|)$ be the total running time of the judgments $\Gamma \vdash f : \_ \rhd \Gamma'$ and $\Gamma' \models \_$ for some $\Gamma'$. 

Building constraints for the typechecking judgment $\Gamma \vdash_\top a : T \rhd \Gamma'$ takes time 
$$\varphi(|a|) = \mathcal O(|a| + \Sigma_{i \in 1..n}\psi(|f_i|))$$ if $a$ contains as subterms the  processes $\mathsf{pack}(f_1),\dots,\mathsf{pack}(f_n)$ without nesting. Checking the satisfiability of those constraints reduces to detecting cycles in a graph, and takes time $\mathcal O(|a|))$, so the total running time for typechecking is 
\begin{eqnarray*}
\Phi(|a|) & = & \varphi(|a|) + \mathcal O(|a|) \\
& = &\mathcal O(|a| + \Sigma_{i \in 1..n}\psi(|f_i|))
\end{eqnarray*}
Next, building the label constraint for the typechecking judgment $\Gamma \vdash f_i : T \rhd \Gamma'$ takes time $\mathcal O(|f_i| + \Sigma_{j \in 1..m_i}\varphi(|a'_{ij}|))$ if $f_i$ contains as subterms the processes $[{\labp}_{i1}]~a'_{i1},\dots,[{\labp}_{im_i}]~a'_{im_i}$ without nesting. Finding $\lab$ such that $\Gamma' \models \lab$ takes time $\mathcal O(\mathbb L|f_i|))$, since at most $\mathcal O(\mathbb L)$ labels need to be checked. So 
\begin{eqnarray*}
\psi(|f_i|) & = & \mathcal O(|f_i| + \Sigma_{j \in 1..m_i}\varphi(|a_{ij}|)) + \mathcal O(\mathbb L|f_i|) \\
& = &\mathcal O(\mathbb L|f_i| + \Sigma_{j \in 1..m_i}\varphi(|a_{ij}|))
\end{eqnarray*}
Plugging the expansion of $\psi$ into the expansion of $\Phi$, and solving by induction: 
\begin{eqnarray*}
\Phi(|a|) \!\!\!& = \!&\! \mathcal O(|a| + \Sigma_{i \in 1..n}\mathbb L|f_i| + \Sigma_{i \in 1..n,j \in 1..m_i}\varphi(|a_{ij}|)) \\
\!\!\!& = \!&\! \mathcal O(\mathbb L|a|)
\end{eqnarray*}

\end{document}